**Shining Light on Photoluminescence Properties of Metal Halide Perovskites**

*Katelyn P. Goetz, Alexander D. Taylor, Fabian Paulus and Yana Vaynzof\**

Dr. Katelyn P Goetz, Dr. Alexander D. Taylor, Dr. Fabian Paulus, Prof. Dr. Yana Vaynzof
Integrated Center for Applied Physics and Photonic Materials (IAPP) and Center for Advancing Electronics Dresden (cfaed), Technical University of Dresden, Nöthnitzer Straße 61, 01187 Dresden, Germany

E-mail: yana.vaynzof@tu-dresden.de

Lead halide perovskites are a remarkable class of materials that have emerged over the past decade as being suitable for application in a broad range of devices, such as solar cells, light-emitting diodes, lasers, transistors, and memory devices, among others. While they are often solution-processed semiconductors deposited at low temperatures, perovskites exhibit properties one would only expect from highly pure inorganic crystals that are grown at high temperatures. This unique phenomenon has resulted in fast-paced progress toward record device performance; unfortunately, the basic science behind the remarkable nature of these materials is still not well understood. This review assesses the current understanding of the photoluminescence (PL) properties of metal halide perovskite materials and highlights key areas that require further research. Furthermore, the need to standardize the methods for characterization of PL in order to improve comparability, reliability and reproducibility of results is emphasized.

# 1. Introduction

The performance of optoelectronic devices based on perovskites has risen dramatically in the past decade.[1] With perovskite solar cells having reached a remarkable 25.2% photoconversion efficiency (PCE),[2] the Shockley-Queisser limit is in sight and it seems like a short matter of time before they surpass benchmarks provided by inorganic semiconductors such as crystalline silicon and gallium arsenide.[3] Part of this rapid rise is certainly due to the knowledge and experience that has been successfully translated from other semiconductors, such as organic polymers and dye-sensitized solar cells.[4] However, the main contributing factor to this success is the natural suitability of the photophysical properties of perovskite materials for thin-film photovoltaics. The high performers are direct band-gap semiconductors with a sharp absorption onset,[5] and their charge carriers show high mobility[6] and long diffusion lengths.[7–9] Their tolerance to defects – with the most probable vacancies/substitutions/interstitials forming within the conduction and valence bands rather than in the energy gap[10,11] – means that fast, low energy-cost processing methods are routes to record performance.[12,13] Compositional tuning is achieved by simple means and translates to bandgap tuning, making perovskites promising for tailor-made electronics.[14–16] Despite these features, some important challenges remain, impeding commercialization.[17] The observation of nonradiative losses indicates that defects cannot be entirely neglected.[18,19] This is certainly the case for aging films and devices, as perovskites degrade under exposure to light,[20] humidity,[21] and oxygen,[22] processes which are in part mediated by defects.[23,24] Ion migration has also been shown to play a strong role in such non-ideal behavior as hysteresis[25–28] and, again, degradation.[29,30] Improvements in film fabrication strategies are needed to solve these problems,[31] but the vast parameter space will require an approach beyond incremental optimization.[32] Beyond these structural and electronic problems, efforts toward photon management through advanced optical design are promising but still in their nascent stages, requiring further research.[33] Advancing the fundamental understanding of light-matter interactions in these remarkable semiconductors will broadly address these deficiencies, allowing for better devices to be designed and engineered.

A simple but versatile tool to achieve this understanding is photoluminescence (PL) spectroscopy. First examined at least as long ago as the mid-19$^{th}$ century, this physical-chemical phenomenon is still seeing new uses in spectroscopic monitoring tools for both commercial products and fundamental discoveries, with applications in fluid dynamics, protein mechanics, health monitoring, and, as here, the development of thin-film semiconductors for optoelectronic devices.[34] At its simplest level, time-averaged, steady state PL requires only the use of a dark room, a continuous wave (cw) laser, and a spectrometer, and allows for examination of relative behavior of luminescing material. Measurement of the PL quantum efficiency (PLQE) requires the addition of an integrating sphere to successfully scatter absorbed light.[35] At the next level of sophistication, short laser pulses and ultrasensitive spectrometers can be used to study the time-dependent phenomenon, allowing for measurement of carrier lifetimes, recombination kinetics, and carrier densities. When combined with microscopy or thin-film fabrication tools, or probed under varying environments or temperatures, it allows for the study of processes *in situ*, making PL spectroscopy indispensable for those wishing to engineer a better optoelectronic device.

As a result of this, PL spectroscopy has been used extensively and in a wide variety of ways for the study of perovskites for optoelectronic applications. Since its optimization coincides with the optimization of the solar cell efficiency under the detailed balance limit,[36] reporting the PLQE is commonplace, especially when concerned with the study and mitigation of defects. At present, however, the same perovskite composition excited by the same power density is reported to have different values for the PLQE depending on the group (**Figure 1**).[37–48] Not only is this suggestive of reproducibility issues which are known to plague perovskites, including varying light outcoupling efficiency and subtle differences in the defect content, but, in light of numerous reports to be discussed in this review, it is suggestive of varying measurement conditions. Therefore, in this review, we discuss the current state of understanding of photoluminescence measurements in perovskite materials for optoelectronic applications with an emphasis on unifying results and understanding across the field. We primarily discuss bulk perovskite thin films and crystals, encompassing a variety of organic, inorganic, and mixed perovskite systems, but touch briefly on quantum dots (QDs) as needed. We begin by outlining the photophysics of the absorption and emission specific to these materials, including a discussion of the measurement practices currently in place (section 2). This is followed by a discussion of chemical and compositional variations, as these, the particular defect chemistry, the processing, and in films the microstructure, each impact the radiative and non-radiative recombination dynamics (section 3). Further study in this area is critical, as the relationship between supposedly defect-tolerant band structures, their microstructure, and the PL lead to non-intuitive and sometimes inconsistent results. Section 4 then highlights instances where photoluminescence is used as a tool to better understand ion migration, a physical phenomenon that has been associated with low device lifetimes and non-ideal device operation (e.g. hysteresis) in solar cells. Once defects are known, they can be passivated in order to maximize the PLQE; this is discussed in section 5. To finalize the discussion of properties relevant to the bulk crystal or film, we discuss the impact of sample history and the measurement environment on the results of PL characterization (section 6). Because the current divergence of methodologies prevents accurate comparisons between research groups, we recommend measurement and reporting parameters which, if adopted by enough researchers, will improve the reliability and utility of the PL measurement.

## 2. Physical Principles of Photoluminescence in Perovskites

Any researcher interested in increasing the efficiency of a light harvesting or generating device to its theoretical maximum must also be interested in understanding and maximizing its luminescence, and perovskites are not exempt from this rule.[49] For light-emitting diodes (LEDs), this is very simple to understand, as the figure of merit is the external luminescence quantum efficiency (ELQE), which is exactly the light observed coming out of the device. Thus, optimizing the radiative recombination of the injected charges and the outcoupling of their emitted photons are the primary engineering tasks when fabricating LEDs. While the three steps governing the operation of a solar cell (light absorption, charge generation, and charge collection) contain no mention of light emission, the radiative efficiencies of the material and device are nonetheless directly and indirectly related to their ability to turn photons into electricity. This is because of its critical relationship to the open-circuit voltage ($V_{OC}$); any non-radiative recombination due to trap states or other parasitic loss pathways leads to both a lower luminescence and a smaller quasi-Fermi level splitting, reducing the maximum possible operating voltage and efficiency.[36,50,51] As such, measuring the steady-state and time-resolved photoluminescence of perovskite films gives important information about its non-ideal material properties, including its defect content and behavior. Here, we briefly discuss the photophysical principles governing the process of PL in perovskites.

## 2.1. Measuring PLQE

There are a variety of ways to use PL, each yielding different sorts of information about the light-matter interactions of the perovskite material. In its simplest form, the steady-state PL is measured by exciting the valence states in the material with a light source and measuring the emission intensity. This is perhaps the most frequently used PL experiment, and is often used to evaluate the luminescence of one sample relative to another, for example given a change in a processing parameter. While qualitative, it can still be used to infer the relative presence of defects (depending on the excitation density). The external PLQE, often abbreviated $\eta_{ext}$, is a quantitative measure defined as the ratio between the number of photons emitted (reaching the spectrometer) to the number of photons absorbed by the sample in question. This is the value of most concern for this review; therefore, we detail the measurement and calculation of this value here. Measurement of $\eta_{ext}$ is most often accomplished by the method of de Mello, Wittmann, and Friend,[35] making use of an integrating sphere to scatter all light except the immediately incident beam, accounting for any angle-dependence of the PL spectra. In this simple but elegant method, four steps are required in order to isolate $\eta_{ext}$, as shown in **Figure 2**. First, a background spectrum (no laser, 2a) is collected to account for any stray light. This spectrum can be immediately subtracted from the other three obtained in the course of the measurement. Second, a measurement of the laser intensity is made without the sample present (b), used to find the absorption coefficient and therefore the number of absorbed photons. Third, the spectra is measured with the sample inside the sphere but not in the laser beam path (c) to account for any reabsorption and reemission of light scattered by the sphere wall. Lastly, the direct excitation of the sample is measured (d).

For the spectrum from each configuration, the integrated intensity for the laser signal is defined as $L$ (e.g. $L_b$, $L_c$, $L_d$) and for the photoemission signal as $P$ (e.g. $P_c$, $P_d$; as there is no sample present $P_b = 0$). We define the fraction of light absorbed by the sample after direct excitation as $A$, as well as a fraction absorbed by the sample after the light was first scattered by the sphere wall $S$. $L_c$ and $L_d$ can be defined in terms of the initial total laser intensity $L_b$, and the absorption coefficients $A$ and $S$. $L_c$ is equal to the initial laser light minus any light absorbed after scattering from the sphere wall (1-S) (**Equation 1**), while $L_d$ is that same term less any light directly absorbed (1-A) by the sample, (**Equation 2**). Substituting Eq. 1 into 2 yields a formula for $A$ in terms of measureable quantities (**Equation 3**).

$$L_c = L_b(1 - S) \qquad (1)$$

$$L_d = L_b(1 - S)(1 - A) \qquad (2)$$

$$A = \left(1 - \frac{L_d}{L_c}\right) \qquad (3)$$

With the value for $A$ known, the value for $\eta_{ext}$ can be derived from the spectrum obtained from configuration (d). This $L_d + P_d$ spectrum has two contributions. The first is the light emitted by the sample, which is equal to the product of the initial laser intensity $L_b$, the direct absorption coefficient $A$, and the external photoluminescence efficiency $\eta_{ext}$ (**Equation 4**, first term). The second is any light that is not directly absorbed by the sample, which is equal to $L_c$ and $P_c$ (Eq. 4, second term). (Note that this second term relies on the assumption that the distribution of light within the sphere is uniform, and that the surface from which light is scattered is unimportant, which the author notes was experimentally verified to within 2%). Eq. 4 can then be solved for $\eta_{ext}$ all in terms of known quantities to give the final result, **Equation 5**.

$$L_d + P_d = AL_b\eta_{ext} + (1 - A)(L_c + P_c) \qquad (4)$$

$$\eta_{ext} = \frac{P_d - (1 - A)P_c}{L_a A} \qquad (5)$$

We encourage researchers to measure the PLQE when feasible to allow for quantitative comparison of spectra and efficiencies between groups.

## 2.2 Recombination in Perovskites

The fundamental definition of the PLQE means that its value is directly related to the extrinsic and intrinsic losses in the sample, which can be described under the general umbrella of non-radiative recombination. Therefore, understanding any details about the luminescence first requires an understanding of charge-carrier recombination. The properties of this have been recently reviewed in detail from several unique perspectives for perovskites;[18,52–56] here we briefly outline what is necessary to motivate and understand the results described in this paper. Recombination can be described by a general rate equation, where the change in charge-carrier density (given by $n$) over time is a combination of three terms, as shown in **Equation 6**[54]:

$$-\frac{dn}{dt} = k_1 n + k_2 n^2 + k_3 n^3 \tag{6}$$

The first term depends linearly on $n$, with $k_1$ being the rate constant for monomolecular carrier recombination. Such recombination is either exciton-mediated or trap-mediated and non-radiative in nature (i.e. Shockley-Read-Hall, SRH, recombination). The quadratic term describes the electron-hole (bimolecular) recombination, which is radiative and proportional to the electron ($e$) and hole ($p$) densities ($n$) as $n_e n_p = n^2$. Finally, the third-order term describes the three-body process of Auger recombination. The expression of each term in a given photoluminescence experiment depends on the value of the photoexcitation density $n$, which is governed by the power of the incident laser and the absorption coefficient of the material. At low photoexcitation densities of approximately $n < 10^{16}$ cm$^{-3}$, $k_1$ dominates; at approximately $10^{16}$-$10^{18}$ cm$^{-3}$ $k_2$ dominates; while at large $n$ Auger recombination is the primary mechanism. (The value for $n$ here was derived from transient absorption measurements on MAPbI$_3$, MAPbBr$_3$, and MAPbI$_{3-x}$Cl$_x$ films, with abbreviations standing for methylammonium lead triiodide with and without chloride and methylammonium lead tribromide perovskites.)[57] An understanding of the role of different recombination regimes can be understood through fluence-dependent measurement of the PLQE.[58] At low intensities, trap states remain unfilled, and non-radiative SRH recombination leads to a low observed PLQE. At higher intensities, the trap states remain populated throughout the measurement, and the radiative regime dominates. This trend continues until, at very high excitation densities, the PLQE remains constant with further increases in laser power due to Auger processes.

In the charge density regimes at play in perovskite solar cells (under 1 Sun light intensity), the monomolecular and bimolecular regimes dominate, while Auger recombination has a negligible impact. The details underlying each of the monomolecular and bimolecular rate constants are intriguing and have immense impact on the properties of the films for devices. In the monomolecular regime, trap states serve as non-radiative recombination centers, shortening the charge-carrier lifetime and reducing the PCE in a device and the PLQE in a film. As alluded to previously, steady-state PL measurements are frequently used at a qualitative level to gain insight into relative trap densities. For more quantitative information, time-resolved photoluminescence (TRPL) is employed. Here, a light source with short pulses excites the sample, and the intensity of the resulting PL signal is measured with respect to time as the sample relaxes back to the ground state. The instrumentation for this measurement is more complex than that for the PLQE measurement, requiring the use of detection methods sensitive to single photons such as gated CCD detection, time-correlated single photon counting (TCSPC), or detection with a streak camera. Photoexcited carriers can undergo this relaxation via the three pathways described in equation 6, and so information from these decay pathways is encoded in the temporal behavior of the TRPL signal. Generally, there is no clear consensus on how to exactly assign the various components to a given TRPL curve, though to simplify this process most often TRPL is measured at low fluences so that nonradiative recombination dominates. In these instances, qualitatively a lengthening of the TRPL signal indicates a reduction in $k_1$, and therefore, a reduction in trap states. In order to extract parameters, typically some combination of exponentials or stretched exponentials must be used to account for a distribution of trapping states with differing characteristics (capture cross section, trapping efficiency, etc.). As we discuss later in this review (section 3.3 and others), the variety of processing methodologies for different perovskites, but especially $MAPbI_3$, has resulted in the observation of a variety of charge-carrier lifetimes and thus a variety of trap densities. Some of these lifetimes are exceptionally long, on the order of 1 μs[9,59], providing some empirical evidence for the benign nature of defects in perovskites. However, these empirical snapshots are not sufficient to describe the role of defects entirely.

Because of the direct relationship between a high degree of radiative recombination and high PCE, efforts to optimize processing focus on minimizing the prevalence of monomolecular recombination, thus maximizing bimolecular recombination. Study of the processes governing $k_2$ have shown that it is an inverse absorption process[60] that is fundamentally well-suited for solar cells, particularly because it has been observed to be non-Langevin in nature[61]. Here, the ratio between $k_2$ and the charge-carrier mobility was found to be approximately four orders of magnitude below the Langevin limit, allowing for the high charge-carrier diffusion lengths necessary for high PCE[55,58]. A further interesting facet of the discussion surrounding $k_2$ is to what degree excitons are responsible for radiative recombination versus free charges. In MAPbI$_3$ the exciton binding energy ($E_B$) has been measured to be approximately equal to $kT$ at room temperature, making the thermal dissociation into free charges a prevalent process[58,62,63]. As such, it has been experimentally determined that free charges are responsible for $k_2$, with exciton populations and effects increasing at low temperatures and high photoexcitation densities[58]. This is of course a feature of the crystal structure and chemical composition, and is further discussed in section 3.

## 2.3. Self-Absorption and Photon Recycling

Ideally, the value of the external PLQE is high. However, even in the case that excitation does result in radiative recombination the emitted photon may not be observed at the spectrometer, making the internal PLQE ($\eta_{int}$) much higher than the external due to factors extrinsic to the bulk material properties. One feature that may cause this is parasitic absorption by the substrate. Another phenomenon that may impact the escape probability of a photon is that of self-absorption leading to photon recycling. Materials with high absorption coefficients, overlapping absorption and emission spectra (a small Stokes-shift), and a high refractive index – including perovskites – can display such a phenomenon, where a photon resulting from the radiative recombination of the initial photoexcitation is reabsorbed repeatedly. This chain of events can continue until the photon escapes from the sample or device, or until the photoexcited charge carriers undergo non-radiative recombination. This has the impact of reducing the apparent bimolecular rate constant ($k_2$) by a factor related to the escape probability ($P_{esc}$), making the effective rate equation that of **Equation 7**.[56,64]

$$-\frac{dn}{dt} = k_1 n + P_{esc} k_2^{int} n^2 + k_3 n^3 \qquad (7)$$

The important question here is at which carrier and trap densities does this effect impact the observed PL, and whether or not this behavior is important to devices. de Quilettes *et al.* simulated the time-resolved PL for semiconductors with and without photon recycling, as shown in **Figure 3**a-d.[56] At high values of $k_1$ (i.e. high trap densities) photon recycling has little impact on the observed PL (Figure 3c,d). For low $k_1$ – here, below $10^5$ s$^{-1}$ – a large positive impact on the carrier lifetime was observed for samples excited to high initial carrier densities (Figure 3b). Given the predictions and observations regarding the low impact of defects on behavior,[9,37,65,66] we can therefore expect that photon recycling must be taken into account in understanding and managing luminescence. Indeed, Pazos-Outón *et al.* directly observed photon recycling in MAPbI$_3$ films in 2016,[64] and further observations have been made in a variety of compositions since.[57,67–70]

Having established the relevant loss factors in the bulk of the semiconductor, if all are properly (quantitatively) accounted for, the internal PLQE ($\eta_{int}$) can be estimated directly from the external luminescence. An expression for $\eta_{int}$ can be derived by accounting for a series of photon recycling events, resulting in a simple relationship to the external PLQE (**Equation 8**)[57,64]:

$$\eta_{int} = \left(\frac{\eta_{esc}}{\eta_{ext}} + (1 - \eta_{esc})\right)^{-1} \qquad (8)$$

Here, $\eta_{esc}$ is the probability of a photon exiting the surface of the sample, which depends upon material properties such as the relative indices of refraction, film roughness, etc., estimated at 12.7% for a glass/perovskite/air sample.[57] As shown in Figure 3h-f, this model was successfully applied by Richter *et al.* to describe the varying losses due to the combination of a narrow escape cone and photon recycling in perovskite films with and without texturing. A slightly different treatment of the optical configuration of the sample, derived from that used for GaAs films[71] and applied to surface-passivated perovskite films,[37] requires knowledge of the absorption coefficient ($\alpha$), refractive index ($n$), and thickness ($d$) of the sample, as well as the loss ($L = 1 -$ Reflectivity) due to the parasitic absorption of the back substrate (**Equation 9**):

$$\eta_{ext} = \frac{\eta_{int}/2n^2}{\eta_{int}/2n^2 + (1 - \eta_{int}) + L/\alpha d} \tag{9}$$

**2.4. From Films to Devices: The Open-Circuit Voltage and Photon Management**

As famously described by Shockley and Queisser, the detailed balance of absorption and emission means that the maximum $V_{OC}$ (and, as a result, the maximum fill factor and PCE) for a semiconductor with a given bandgap ($E_g$) is observed in the limit where only radiative recombination is present in the sample.[36] The impact this has on efficiency can be understood by considering that the magnitude of the $V_{OC}$ is determined by the photoexcited charge-carrier density. By definition, charges are not extracted at the electrodes (i.e. $J = 0$ mA) at $V_{OC}$, meaning they must recombine. Any defects present within the energetic gap – whether they are intrinsic to the bulk of the semiconductor, its surface, or its interface with extraction layers – result in non-radiative recombination at these energetic states, reducing the quasi-Fermi level splitting ($\Delta\mu = E_F^{CB} - E_F^{VB}$, with the terms $E_F$ defining the effective conduction and valence band Fermi levels) and therefore reducing the charge density. $V_{OC}$ and $\Delta\mu$ are related very simply through the charge of the electron ($q$) as shown in **Equation 10:**

$$qV_{OC} = \Delta\mu \tag{10}$$

Critically, the value for $\Delta\mu$ can be quantitatively related to the PL intensity at photon energy $\varepsilon$ as shown in **Equation 11**:

$$I_{PL}(\varepsilon) = \left(\frac{2\pi\varepsilon^2}{h^3 c^2}\right)\left(\frac{1}{e^{\frac{\varepsilon-\Delta\mu}{k_B T}} - 1}\right) a(\varepsilon, E_F^{CB}, E_F^{VB}, T) \qquad (11)$$

Here, $k_B$ is the Boltzmann constant, $T$ is the temperature, $h$ is Plank's constant, $c$ is the speed of light, and $a$ is the spectral absorptivity (including any sub-gap states)[72,73].

This relationship makes it possible to understand how the PL of the semiconductor layer might result in a particular $V_{OC}$; specifically, a large measured value of $\Delta\mu$ can be used to infer a large $V_{OC}$. However, many studies focusing on measuring the PL of perovskite samples often omit charge extraction layers from their stacks, as extraction layers can significantly quench the PL signal. This can lead to a discrepancy between PL and full PV device results, as parasitic non-radiative absorption in a device is not limited to the bulk of the semiconductor. However, the quality of the interfaces must also be considered. Because the maximum radiative efficiency in the device structure corresponds to a maximum $V_{OC}$, the detailed balance limit also explains that good solar cell must also be a good LED; therefore, injecting charges into a solar cell at $V_{OC}$ allows for a direct relationship between this and the quantum efficiency of the LED ($\eta_{LED}$), as shown in **Equation 12**.[19,50,51]

$$V_{OC} = V_{OC,max} - \frac{k_B T}{q} \ln(\eta_{LED}) \qquad (12)$$

Here, $k_B$ is Boltzmann's constant, $q$ is the electron charge, $T$ is the temperature, and $V_{OC,max}$ is the maximum possible open-circuit voltage after unavoidable thermodynamic and entropic losses.

This concept is already in use to understand the relationship between processing, defects, luminescence, and the operating voltage of perovskite solar cells, with insightful discussions and examples existing in the literature.[19,74,74–78] In a remarkable recent feat, Liu and coworkers fabricated solar cells based on MAPbI$_{3-x}$Cl$_x$ with PTAA and PCBM (respectively, poly(triarylamine) and [6,6]-phenyl-C$_{61}$-butyric acid methyl ester) hole and electron transport layers, achieving repeatably high $V_{OC}$ values of 1.24 V to over 1.26 V[79]; this is within 60 mV of the thermodynamic limit of approximately 1.32 V for MAPbI$_3$. They verified their findings by evaluating the ELQE, the external PLQE and $\Delta\mu$, finding via equations 11 and 12 that they matched expectations. This example demonstrates nicely how luminescence and device properties are related, lending credence to the studies which study surface passivation in a glass/film/passivator configuration (but not in a full solar cell), as later described in Section 5.2. The previous study reporting high $V_{OC}$ in MAPbI$_3$ concluded that the low losses are due to extremely low recombination at the interfaces between the perovskite/extraction layers/contacts.

The reciprocity described in equation 12 can also be put to use to engineer better optical interfaces. This was the case for gallium arsenide (GaAs) solar cells, where an almost perfect internal PLQE of 99.7% was achieved as early as 1993, implying that most of the bulk material problems (e.g. defects) had been solved.[71] Recent devices therefore received a boost in the $V_{OC}$ not due to any improvements in the active layer, but due to the addition of a highly reflective mirror at the back side of the substrate, which served to properly manage re-emitted photons by preventing parasitic absorption at the back substrate.[49] Many similarities have been drawn between the electronic structure and photophysics of perovskites and GaAs, including the direct band-gap electronic structure, efficient luminescence, the impact of photon recycling, and more, leading to discussions as to whether similar photon management may prove effective in minimizing $V_{OC}$ losses in the former material.[33] Though in its nascent stages, addressing the optical design of perovskite surfaces and interfaces is already theoretically and experimentally demonstrated to have a positive effect.[33] As shown in Figure 3g, textured substrate and semiconductor surfaces (Figure 3h) enhance the escape probability of a photon (the high refractive index of the perovskite means it has a narrow escape cone at its air interface), bringing the external PLQE closer to the internal PLQE.[57] Ultrahigh internal and external luminescence efficiencies could be observed for surface-passivated films on highly reflective substrates.[37]

Theoretically, photon recycling has been predicted to enhance the $V_{OC}$ of perovskite solar cells by 50-80 mV, making its impact non-trivial.[52,80] Photon management is also noted to be highly necessary for the design and optimization of functional tandem solar cells, which may prove to be one of the more economically feasible applications of perovskites.[33,81] Currently, however, the problem of defects in perovskites is still not fully solved, making reduction of trap-assisted recombination important in the development of higher efficiency solar cells.[82] Thus, the next section discusses the impact of the perovskite composition on photoluminescence, ranging from the bulk crystal, to defects, to the microstructure of the film.

**3. Composition, Defects, and Microstructure and their Impact on Photoluminescence**

The emission spectrum of a perovskite is intimately related to its electronic structure, which is in turn governed by its chemical composition and crystal structure. Perovskites form in the $ABX_3$ structure of the original calcium titanate, with $A^+$ being a monovalent cation, often cesium, methylammonium ($MA^+$), or formamidimium ($FA^+$), $B^{2+}$ being a divalent heavy metal ion such as lead or tin, and the anion $X^-$ being a halide such as chloride, bromide, or iodide, as shown in **Figure 4**. So long as the Goldschmidt tolerance factor is maintained, whole exchanges at the A, B, and X sites allow for the formation of a stable 3D perovskite.[83,84] These changes impact the electronic structure of the film in different ways, with the halide having the strongest impact on the value of the bandgap. Perhaps most enticing for optoelectronic applications is the potential for halide mixing, where control over the I:Br or Br:Cl ratio in a thin film allows for precise tuning of the bandgap energy, and thus precise control over the emission wavelength.[14] Such compositional tuning, as well as variations due to temperature (including phase changes), yield a vast space for the understanding of structure-function relationships in perovskites, and are the topic of the first part of this section.

Of course, every crystal that exists outside a computer has a non-zero density of imperfections, requiring their discussion in conjunction with the bulk crystal structure in order to understand the emissive behavior of a semiconductor. Somewhat astonishingly, perovskites seem to tolerate a higher level of defects than other semiconductors for solar harvesting,[65,66] a feature that enables such 'quick and dirty', energy and cost-effective fabrication methods by a range of solution processing techniques. The high PCEs observed in solar cells verify this to a degree, with theoretical calculations in at least partial agreement;[66,85] however, the observation of non-unity internal PLQE, among other measurements, indicates that defects still do play a role in the observed emission and related phenomena, which will be discussed in the second part of this section. Finally, we survey the effect of film microstructure on its photoluminescence properties and discuss the observations of spatial PL inhomogeneities and their effect on device performance.

## 3.1. ABX Combinatorics

*3.1.1. A-Site Modifications*

The MAPbI$_3$ perovskite, containing methylammonium, lead, and iodide, is the most commonly studied material composition, and is often used as a point of comparison for other structures. MAPbI$_3$ thin films exhibit sharp PL spectra centered at ~775 nm,[86] consistent with a band gap of approximately 1.6 eV, with quantum dots being somewhat blue shifted due to quantum confinement effects, though this effect is smaller for perovskite QDs than for other material systems.[87,88] At the A-site is the organic cation MA$^+$, with an effective ionic radius of ~220 pm. For a different cation in this position to form a 3D perovskite, which is often the phase of optimal photoactivity, it must be of the correct size in order to slot into the perovskite crystal structure ABX$_3$. This size is given by the Goldschmidt tolerance factor (**Equation 13**), denoted as $\alpha$:

$$\alpha = \frac{r_A + r_X}{\sqrt{2} \cdot (r_B + r_X)} \tag{13}$$

where $r_i$ are the ionic radii of each of the proposed components.[89] To form the desired black phase perovskite for this composition, $\alpha$ must lie between 0.8 and 1.0, as shown in Figure 4c[90]. Perhaps surprisingly, this condition dramatically restricts the possible A-site substituents to three candidates: $Cs^+$, $MA^+$, and $FA^+$. In addition, molecules which are "close enough" to the desired size, such as the too-small rubidium[74] or the too-large guanidinium,[91] can be successfully incorporated in smaller amounts.

At first glance, changing the material in this site should only have a limited impact on the band gap, and thus the emission wavelength, as the states which make up the conduction and valence band frontiers are located on the $BX_6^{4-}$ octahedron, specifically in an antibonding state between the B $n$s and X $m$p orbitals.[92] A change in the ionic radius at the A-site, though, indirectly causes a change in the band gap energy by inducing steric strain on the crystal lattice, expanding or contracting the lattice and altering the overlap between the frontier orbitals. Replacing $MA^+$ with $Cs^+$ in the A-site causes a blue shift in the PL signal due to the smaller effective ionic radius (~170pm),[93] leading to a larger overlap between the Pb and I orbitals and thus a higher energy state. Unfortunately, early investigations found that the desired photoactive "black" phase for pure cesium lead triiodide ($CsPbI_3$) perovskite is unstable at temperatures under ~310°C.[94] By employing solvent engineering techniques the "black" phase could be stabilized, leading to a PL signal for pure $CsPbI_3$ at around 700-725 nm,[95] consistent with a bandgap of 1.73 eV.[96] Luo and coworkers went a step further, and used the quenching of the PL signal for the perovskite fabricated on a $TiO_2$ underlayer as an indicator of efficient electron extraction, for potential application in solar cells.[97] In the other direction, substituting in the larger cation $FA^+$ (effective ionic radius ~250pm[93]) expands the crystal lattice, forcing apart the Pb and I orbitals and lowering the band gap energy to approximately 1.48 eV[98] and a PL signal centered at 840 nm. As this is closer to the solar cell optimum bandgap of 1.33 eV,[36] there has been significant research on perovskite solar cells incorporating $FA^+$. Again, however, a less stable photoactive phase leads to inferior performance for the pure formamidinium lead triiodide ($FAPbI_3$),[99] and so most focus is instead on so-called alloyed perovskites, consisting primarily of $FA^+$, but utilizing $MA^+$ and/or $Cs^+$ to stabilize the photoactive phase.[100]

It is worth mentioning that the simple model presented above, whereby the ionic radius of the A-site constituent solely determines the band gap, may be incomplete. Using first principle density functional theory (DFT) calculations, Amat and coworkers suggested that the decrease in band gap energy for $FA^+$ containing perovskites is instead largely due to the increased formation of hydrogen bonds between the $FA^+$ and the $PbI_6^{4-}$ octahedral.[101] This hydrogen bonding induces a tilt in the octahedral orientation which enhances the Pb character of the conduction band minimum, increasing the ionicity of the Pb-I bond and therefore amplifying spin-orbit coupling.[102] This result highlights that, while the experimental trends may be simple, the exact physical mechanism responsible for it may be more complex and requires further study.

*3.1.2. B-Site Modifications*

While lead is the largest toxicity risk amongst the perovskite components,[103] currently, lead-based perovskites outperform their lead-free counterparts by a significant margin in both light-generating and light-harvesting applications.[4] Therefore, although high performance lead-free perovskites are highly desired,[104] inferior performance and more difficult processing has led to few reports on the subject. In the cases where they do occur, most substitute the lead with tin (Sn), although other metals such as germanium (Ge) are possible.[105] Performing this substitution leads to a significant lowering of the band gap energy, with the methylammonium tin triiodide ($MASnI_3$) composition possessing a PL peak centered at 950 nm.[106] Interestingly, these materials display a non-monotonic band gap trend for Pb/Sn alloys – instead of increasing as the Pb:Sn ratio increases, as expected, the band gap decreases, for up to ratios of 4:6. This leads to PL spectra centered at 980 and 1000 nm, for 4:6 and 2:8 Pb:Sn ratios, respectively.[107] Another promising avenue for a lead-free perovskite is the so called "double-perovskite," substituting a mono- and trivalent molecule for a pair of the divalent lead molecules, forming the general structure $A_2B'^{+}B''^{3+}X_6$.[108] Among the possible combinations, cesium silver bismuth halide ($Cs_2AgBiX_6$) perovskites have attracted the most attention. For example, dicesium silver bismuth hexabromide ($Cs_2AgBiBr_6$) has been shown to be an indirect bandgap semiconductor ($E_g$=1.95 eV) with a relatively long PL lifetime, promising for optoelectronic applications.[109] However, the observation of a broad PL spectra, low PLQE, and relatively poor photovoltaic performance suggests that recombination in this material is predominantly non-radiative.[110] A recent study by Zelewski *et al*. revealed that PL in this material occurs via color centers rather than band-to band transition.[111] The composition dicesium silver bismuth hexachloride ($Cs_2AgBiCl_6$) has also been realized, exhibiting a larger bandgap in the range of 2.3-2.5eV and a broad PL emission centered at ~550 nm.[112] While dicesium silver bismuth hexaiodide ($Cs_2AgBiI_6$) has not been realized in this bulk form, it has been recently demonstrated that $Cs_2AgBiI_6$ nanocrystals can be synthesized either by anion exchange from $Cs_2AgBiBr_6$ nanocrystals[113] or directly by the choice of appropriate precursors.[114] The iodide-based nanocrystals showed the desired decrease in bandgap ($E_g$ = 1.75 eV); however, their PL was significantly reduced, highlighting the dominant sub-bandgap trapping processes in nanocrystals based on double perovskites.[114] Developing mitigation strategies to increase the PLQE of such structures is of critical importance for their future application in lead-free optoelectronic devices. One such strategy was realized by Nandha *et al*. for the case of the wide bandgap dicesium silver indium hexachloride ($Cs_2AgInCl_6$) nanocrystals.[115] The authors demonstrated that by doping the nanocrystals with manganese cations ($Mn^{+2}$), a moderate enhancement of the PL is possible. A more striking example was recently reported by Luo and coworkers,[116] who cleverly exploited the presence of ultrafast-created self-trapped excitons (STE) in systems of $Cs_2AgInCl_6$ quantum dots to achieve dramatic enhancements in PLQE, from <0.1% up to 86% ($\lambda_{exc}$ = 365 nm, ~25 W cm$^{-2}$). Normally, the excitonic recombination is a "dark" transition (no emission) as the radiative transition is parity-forbidden, leading to the extremely low initial PLQE value. By alloying sodium (Na$^+$) into the B-site, however, the inversion symmetry of the lattice is broken, allowing the previously forbidden radiative recombination of the STE, and stable, efficient white-light emission. This example illustrates the tremendous potential of double perovskite nanocrystals for optoelectronic applications; however with the field being in its infancy, much remains unknown about the fundamentals of recombination processes in these materials, and further research is required to fully utilize their potential in functional devices.[117]

*3.1.3. X-Site Modifications*

Because the conduction and valance band states are located on the $BX_6^{4-}$ octahedra, substitutions on the X site have a dramatic impact on the band gap. Correspondingly, the most commonly reported anions used to substitute for I$^-$ are Br$^-$ and Cl$^-$. The electronegativity increases from I to Br to Cl, thereby increasing the optical transition energy and blue shifting the resulting PL spectra. For example, a full replacement of the I$^-$ by Br$^-$ in methylammonium lead trihalide ($MAPbX_3$) perovskites increases the band gap energy and shortens the emission wavelength to 2.4 eV and 550 nm, respectively, while substituting with Cl$^-$ is a further modification to 3.2 eV and 410 nm, respectively.[118] We note that these substitutions affect not only the bandgap of the materials, but also their exciton binding energies. For example, in the case of $MAPbBr_3$, while literature reports show little agreement about the exact value of the exciton binding energy,[119–121] it is generally accepted that it is larger than that of $MAPbI_3$. The situation is similar in the case of methylammonium lead trichloride ($MAPbCl_3$).[122] The consequence of high binding energies is a more complex photophysics, considering that photoexcitation in such materials leads to the formation of both excitons and free charge carriers. Because their relative densities depend on the excitation power, this can be directly monitored by the relative ratio of the dual emission observed in PL experiments.[123] Despite the wealth of optical studies performed on Br$^-$ and Cl$^-$ based perovskites, an accurate understanding of their photophysics is still lacking and requires further research.

*3.1.4. Mixed-Composition Perovskites and Incremental Tuning*

The combination of all these potential substitutions gives a possible emission range for perovskites of around 600 nm. While this in and of itself is impressive, what really makes perovskites intriguing is the possibility for mixed compositions, where partial substitutions at any of A, B, and X sites can give rise to band gaps (and therefore emission peaks) centered at any point within this range of values. Indeed, many of the most promising and cutting-edge results employ these mixed composition perovskites. For example, in the commonly employed system of cesium lead trihalide (CsPbX$_3$) perovskite quantum dots, the emission spectrum can be tuned across the entire visible color gamut by adjusting the halide ratio, as shown in **Figure 5.**[124] A full list of publications demonstrating this is beyond the scope of this review, but the interested reader can find more information in several reviews on perovskite LEDs.[125–127]

For light-emitting devices, the advantages of tuning the emission wavelength are obvious for color control. In light harvesting applications, higher band gaps are useful in tandem solar cells: using two active layer materials, perhaps silicon-perovskite[128,129] or perovskite-perovskite,[130] with different band gaps allows for the absorption of a wider range of the solar spectrum and therefore a higher maximum efficiency. For use in silicon-perovskite tandem cells, a perovskite absorber with a band gap of 1.75 eV would provide the highest theoretical efficiency due to current-matching considerations with a 1.1 eV band gap silicon absorber.[131]

Achieving this band gap is easily possible using iodide-bromide mixed perovskites, however when researchers characterized these materials, contrary to what is expected for a higher band gap PV cell, they found that the $V_{OC}$ did not improve, or sometimes even worsened, when compared to pure MAPbI$_3$.[14,132] Seeking to understand this effect, Hoke *et al*. measured the PL spectra for films of methylammonium lead mixed halide MAPbBr$_x$I$_{3-x}$ under constant illumination and found that over time a second PL peak emerged at higher energy.[133] With additional characterization, they concluded that under illumination the halide ions redistribute within the film, separating into iodide- and bromide-rich regions, which results in lower quasi-Fermi level splitting and the lower observed $V_{OC}$. A range of studies demonstrated that such light -induced phase segregation can revert back to the original MAPbBr$_x$I$_{3-x}$ phase in the dark.[134–136] Similar observations, albeit to a far lesser degree, have been reported for all-inorganic mixed halide perovskites (CsPbBr$_x$I$_{3-x}$)[137,138] and mixed cation mixed-halide perovskites.[139,140] The use of PL to observe the phenomenon of ion migration in perovskite films will be further discussed in Section 4.

It is of course possible to have perovskites with both mixed cation (Rb$^+$/Cs$^+$/MA$^+$/FA$^+$) and anion (I$^-$/Br$^-$/Cl$^-$) together, one example being the high solar cell performance "quadruple cation" composition (RbCs)$_x$(MA$_{0.15}$FA$_{0.85}$)$_{1-x}$Pb(I$_{0.85}$Br$_{0.15}$)$_3$, which yields a PL signal centered at 770 nm.[74] Here, the competing effects on the PL spectra of the primary cation FA$^+$, and minority anion Br, essentially cancel each other out, yielding a slightly shorter emission wavelength as compared to pure MAPbI$_3$. A good rule of thumb for approximating the relative magnitude of competing changes is that modifications to the halide position are about 7x more effective than the cation position at adjusting the band gap energy.[14] Thus, while replacing MA$^+$ with FA$^+$ decreases the band gap energy by 0.1 eV, the ~15% replacement of I$^-$ by Br$^-$ will increase it by 0.15×7×0.1 eV = 0.105 eV, resulting in a largely unchanged band gap.

**3.2. The Impact of the Crystal Structure and Temperature on Photoluminescence**

Most studies on perovskite devices are conducted in a controlled environment and at room temperature, however devices under operation are necessarily exposed to the elements and subjected to a wide range of temperatures. In fact, a realistic consideration of the operating conditions for a PV panel concludes that working temperatures could easily exceed 325K.[141] Therefore, an understanding of the physics of perovskite films as a function of changing temperature is critical to a real-world implementation of devices. Temperature-dependent PL is one way with which to directly probe the photophysics of perovskite films.

*3.2.1. Measurement of the Exciton Binding Energy*

As mentioned previously, an important question to the operation of both LEDs and photovoltaics (PVs) is whether excitations are described by free charge-carriers or excitons. Determination of the exciton binding energy ($E_b$) relative to the surrounding temperature lends insight into this question. This value, of course, is highly dependent on the chemical composition of the perovskite in question. Temperature-dependent PL can be used to estimate the value of $E_b$ by integrating the total PL intensity and plotting it as a function of temperature, as shown in **Figure 6**. The first observation is that, counter to what is commonly observed in other semiconducting materials such as GaAs, the PL peak blue-shifts with increasing temperature, caused by the reverse band edge ordering in perovskites.[102] Second, as temperature increases from the theoretical minimum 0 K the PL intensity decreases, as the additional thermal energy increases the probability that the exciton will dissociate according to a Boltzmann distribution[142] (**Equation 14**),

$$I(T) = \frac{I_o}{1 + Ae^{(-E_b/k_bT)}} \tag{14}$$

where $I_o$ is the luminescence at 0 K, $k_b$ is the Boltzmann constant, and A is a fitting parameter. Using this method, values for $E_b$ were estimated to be between 19 and 32 meV for MAPbI$_3$,[143,144] and 62 meV for methylammonium lead mixed halide (MAPbI$_{3-x}$Cl$_x$)[145] films.

For the bromide compositions MAPbBr$_3$ and formamidinium lead tribromide (FAPbBr$_3$) these values were similar to those with iodide – 53 and 22 meV, respectively.[146] This wide range of values is not that surprising given the complexities of radiative emission in perovskite films and their sensitivity to exact fabrication methods and conditions,[55] and is considered more so an upper bound rather than exact value.[142] However, a key result is that $E_b$ for some perovskite compositions is below or comparable to the thermal energy at room temperature of ~26 meV, leading to the conclusion that the charge-carrier dynamics of devices using these films are dominated by free charges rather than excitons, while others show higher $E_b$ values, suggesting that both species co-exist upon photoexcitation of these perovskites.[121,123] With the broad range of exciton binding energies reported for the various perovskite compositions, significantly more research is required in order to elucidate the exact nature of their photophysical properties.

*3.2.2. Crystal Structure and Phase Transitions*

Turning our attention away from the temperature-dependent PL intensity, the temperature-dependent PL peak position and width yield additional information. Phenomena important to the operation of perovskite devices, such as phonon interactions, lattice dilation, and charge-carrier trapping can all be derived from these features.[147–149] in addition to the exciton behavior discussed above. Furthermore, using other methods, such as X-ray diffraction and thermogravimetric analysis,[14,98,150] several crystal phase transitions have been identified for a wide number of perovskite compositions, which can significantly impact the observed PL signal.[150–152] While the phase-change behavior of each perovskite composition is unique, for lead-based perovskites generally a transition from orthorhombic to tetragonal occurs at approximately 140-160K, with another composition-dependent transition occurring at higher temperatures. Thus, the temperature dependence of the PL can give useful information about the detailed interplay between structural aspects of the perovskite and its optoelectronic behavior.

One of the first reports of temperature-dependent PL that covered a wide range of temperatures was performed by Milot *et al.* in 2015.[148] Here, using a variety of analysis techniques, the researchers examined the charge carrier dynamics of MAPbI$_3$ thin films in detail. Temperature-dependent PL measurements revealed the same blue-shift trend with increasing temperature as seen before; however, they also found two red-shifting discontinuities located at 160 and 310 K. These correspond to the known MAPbI$_3$ phase transitions: from orthorhombic to tetragonal, and from tetragonal to cubic. The size of the PL shift was found to be significantly larger for the 160 K transition, consistent with the orthorhombic phase having a larger $E_g$.[153] This is due to the MA$^+$ ions being highly ordered, or "locked" in the orthorhombic phase,[154] thus generating a significant electric field and increasing $E_g$ by the Stark effect. In the tetragonal and cubic phases, the MA$^+$ ions are "unlocked", and free to rotate about, resulting in a smaller bandgap and easier exciton dissociation as the MA$^+$ ions can screen the electric field.[155] In addition, the authors found that for low temperatures (< 120 K), a second PL peak is observed at higher energies, increasing in intensity with decreasing temperature until it becomes dominant at very low temperature (8 K). This dual emission has also been observed for MAPbI$_x$Cl$_{3-x}$[145] and MAPbBr$_3$,[146,156] but not for formamidinium lead trihalide FAPbX$_3$[157] or CsPbX$_3$[158] perovskites, indicating that MA$^+$ plays a key role in this feature.

Seeking to explain this effect, Baikie *et al.*[159] and Wehrenfennig *et al.*[160] proposed that a gradual transition from tetragonal to orthorhombic is impossible, and so small inclusions of the lower band gap tetragonal phase remain present after the phase transition, leading to two PL emission sources within the film. However, Dar *et al.*[161] pointed out that this explanation fails to account for the absence of dual emission from FA$^+$ containing perovskites. Instead, using DFT and molecular dynamics (MD) simulations in concert with PL spectroscopy, they proposed that in certain film regions the MA$^+$ molecules in the tetragonal phase are "frozen" in their disordered state during the phase transition. These randomly oriented MA$^+$ ions' electric field contributions then cancel out and no longer modify by the band gap via the Stark effect. Lastly, Wright *et al*.[157] examined the temperature dependent PL peak position and width for four commonly employed perovskite compositions, MAPbI$_3$, MAPbBr$_3$, FAPbI$_3$, and FAPbBr$_3$, the results of which are displayed in color maps in **Figure 7**. The previously discussed blue shifts with increasing temperature and dual/broadened emission for MA$^+$ containing perovskites at low temperatures are all present.

Seeking to understand the charge-phonon interactions in perovskite materials, the authors analyzed the temperature dependent full width half maximum (FWHM) of the PL signal for each composition studied.[162] This relationship is critical for electronic devices, as charge carrier-phonon coupling sets the theoretical maximum for charge carrier mobility.[163] By fitting the experimentally measured FWHM for each material to the temperature dependent PL linewidth $\Gamma(T)$, which is a sum of the various coupling contributions shown in **Equation 15**:

$$\Gamma(T) = \Gamma_o + \Gamma_{ac} + \Gamma_{LO} + \Gamma_{imp} \tag{15}$$

where $\Gamma_0$ is the temperature independent broadening and is a consequence of lattice defects, $\Gamma_{ac}$ and $\Gamma_{LO}$ are the contributions from scattering off of acoustic and optical phonons, and $\Gamma_{imp}$ is the contribution from scattering off of ionized impurities within the crystal lattice. To see the detailed form for each term in Equation 15, refer to references 161 and 162. Doing so revealed all four compositions possessed high quality electronic band structures, with negligible contributions from defects or impurities. Instead, the dominant contribution to the FWHM broadening came from the Fröhlich coupling of charge carriers with the longitudinal phonons (LO-phonons). Furthermore, the effect was approximately 50% higher for Br$^-$ containing perovskites compared to I$^-$ containing perovskites, suggesting lower charge carrier mobilities for these compositions, as well as confirming that material composition can impact the mobility.

It is important to note, however, that other research has confirmed that perovskites are coupled electronic-ionic systems, and so a description of transport in perovskites would be incomplete without considering ionic motion. A detailed description of the measurement of this ion migration in perovskites will be given in Section 4.

### 3.3. Defects and Photoluminescence

Energetic states lying deep in the bandgap serve as centers for non-radiative recombination, reducing the observed emission in the thin films and devices, and additionally resulting in $V_{OC}$ losses for solar cells. As such, increasing densities of deep trap states reduce the PLQE primarily by contributing to an increase in the amount of monomolecular relative to bimolecular recombination. In the steady-state PL, this manifests as a reduction in the peak intensity, while in time-resolved PL measurements, a change in the shape of the transient is observed, and the PL lifetime is shorter. For perovskites, defect densities in thin films have been measured at a level of $10^{14}$-$10^{18}$ cm$^{-3}$.[164–166] Intriguingly, these values are more comparable to those observed in high quality organic semiconductors[167–169] than materials that demonstrate PCEs over 20%, such as crystalline silicon, which displays a defect density of $10^8$ cm$^{-3}$ at its lowest.[170,171] The remarkable PCEs observed in perovskites in spite of these high densities have sparked excitement over their so-called defect tolerance,[11] the origin(s) and exact meaning of which are still under debate[172,173]. Early on, Yin *et al.* performed density functional theory (DFT) calculations, showing that the defects with a high probability of formation in MAPbI$_3$ lie only at shallow levels[66]. This was discussed to be a result of the valence band extremum being defined by the antibonding coupling between the Pb lone-pair *s* orbital and the I *p* orbital, as well as the high iconicity of the perovskite. Other groups have independently predicted the shallow nature of the most probable point defects.[66,85,174–176] Long carrier lifetime has also been attributed a carrier protection mechanism caused by high dielectric screening and large polaron formation[177–179]. In contrast to these ideas, ongoing research suggests that defects cannot be neglected, and the theoretical and experimental picture is far from unified.[172] The observation of defect-dependent external PLQE,[180–182] as well as deep level transient spectroscopy (DLTS) measurements, which probe in-gap states through capacitive transients, suggests that deep trap states do appear in non-trivial quantities under certain processing conditions.[41,164,180–182] Theoretical arguments that take into account processing-related effects, such as a halide-rich or poor environment, also indicate that the defect formation energy is impacted by the processing environment.[175] Here, we discuss the impact of specific defects (where they are known) on the measured value of the PLQE. While some defects have a conclusively negative impact on luminescence and device efficiency, others are more positive, pointing toward the potential for defect engineering in perovskites.

Single crystals can be a particularly nice platform to understand the fundamental impact of defects on the emission because their absence of microstructure means that the effects of grain boundaries and orientational disorder do not need to be considered. Kim *et al*. measured the facet-dependent emission ($\lambda_{exc}$ = 409 nm, one-photon excitation) on MAPbI$_3$ crystals grown by the inverse temperature crystallization process,[183] showing a strong anisotropy of properties, which they attributed to native point defects.[184] They found that the emission intensity from the (112) crystallographic facet is reduced and red-shifted when compared to the (100) face; these occur at 784 nm and 776 nm respectively. In order to understand how defects play a role in this effect they measured the contact potential difference (CPD) of the two facets by scanning kelvin probe force microscopy. Illumination of the sample induces a photovoltage; by attributing positive shifts to a more positively charge surface and vice versa, they can understand which type of majority charge carrier is present at each surface as they proceed through the experiment. Reasoning through the sorts of ion migration that occurs during their measurement process, they find that the (100) facet displays n-type behavior and suspect that iodide vacancies are dominant. The (112) facet, on the other hand, displays p-type behavior, suspecting that the dominate defects are $MA^+$ and $Pb^{2+}$ vacancies, which create shallow traps near the valence band, while also allowing for the possibility of the coexistence of $MA^+$ and $Pb^{2+}$ interstitials. These latter gaps are predicted to lie mid-gap, which would explain the reduced PL efficiency.

A detailed, measurement-based understanding of the impact of defects on the photoluminescence of thin films and associated phenomena is rather difficult to achieve. Any changes to the processing parameters which might serve to modify the defect content are also likely to alter other aspects of the film, particularly the microstructure, grain size, and crystal orientation, all of which will also impact the optoelectronic properties of the film (as we discuss in detail in the following section). Nonetheless, a few studies have managed to convincingly incorporate defects into the films and show their effects. In particular, crystal-growth environments which are halide-deficient are predicted to have defects with different energies of formation than those with excess halide, as shown in **Figure 8**a-d.[175] Fassl *et al.* varied the density of surface defects in $MAPbI_3$ films derived from the one-step lead acetate trihydrate ($PbAc_2 \cdot 3H_2O$) recipe[185] by incrementally increasing ratio *y* of the iodide-containing MAI precursor to the lead-containing precursor in solution by $\Delta y = 0.01$-$0.02$, ranging from 2.96 to 3.06[41,180]. Such ratios are outside the ideal *y* = 3 by only very small amounts and represent errors that could occur unintentionally during film fabrication. The variation in precursor stoichiometry was connected to the defect content via X-ray photoemission spectroscopy, which, while unable to evaluate the precise chemical nature of the defect does show a change in the ratio of iodine to lead at the surface, which directly (linearly) corresponds to the precursor changes.[41] The impact on the value of the PLQE is rather large:[41,180] iodide-deficient films showed much higher external PLQE, reaching almost 9% at the outset of the measurement, while iodide-rich films showed negligible external PLQE ($\lambda_{exc}$ = 532 nm, ~2 sun, dry $N_2$). These values are shown in Figure 8e. Microstructural changes to the films were negligible, allowing varying grain size to be ruled out as a source of an increased surface defect quantity, nor was the absorption onset affected. The energetic disorder in the film, represented by the Urbach energy, was, however, found to increase as a function of increasing iodide content, though all values were small: under 22 meV. Notably, when incorporated into devices, the $V_{OC}$ of the iodide-deficient films was lowest, increasing almost linearly as the iodide content was incrementally increased.[41] This is attributed in part to an increase in the built-in potential of the device caused by the variation in the energetic structure of the perovskite surface by these defects. Similar experiments on MAPbBr$_3$ films ($\lambda_{exc}$ = 405 nm, ~1 sun, dry N$_2$) showed the same trend in external PLQE (Figure 8e), indicating that not only does the iodide chemistry induce non-radiative recombination centers in films, but so also does the bromide chemistry.[181] A key difference between this study and the iodide study is that the microstructure does change as a function of the precursor solution composition, preventing its complete decoupling from the observed PLQE behavior. The impact of the iodide also has an effect when incorporated in a sequential deposition route. Kong, *et al*, first fabricated lead iodide films, and then immersed these in an MAI-dimethylcarbinol (DMC) solution for 3 minutes.[186] To control the iodide content in the films, one set was washed with excess DMC to remove residual MAI at the surface, creating an iodide-poor sample. The other set was removed from the MAI solution and dried without washing, creating an iodide-rich sample. The detrimental effects of the excess iodide on trap state formation is observed in the PL spectra (Figure 8f,g) as both a reduction in the steady-state PL and the in the carrier lifetime.

Recently, Nan *et al*.[155] examined the impact of small amounts of chloride on the electronic structure of MAPbI$_3$ *via* time-dependent density functional theory (DFT), finding that it results in the transition from defect-localized to extended excited states. This effect is qualitatively summarized in **Figure 9**a. To verify this experimentally, they created films of MAPbI$_{3-x}$Cl$_x$ with x varying incrementally from 0 to 0.1, finding that the PL behavior supports this trend. As shown in Figure 9b, the samples with higher chloride content show a higher steady state intensity, to an extent; the PLQE maximizes at 2.3% for the samples with x = 0.05 ($\lambda_{exc}$ = 532 nm, intensity unknown, atmosphere unknown), with increasing chloride also being accompanied by a red-shift in the peak position. Time-resolved PL measurements demonstrate that the increase in PLQE is accompanied by a decrease in the trap density, found to be on the order of $10^{15}$-$10^{16}$ cm$^{-3}$ (as reported in the text). Note that this occurs in addition to microstructural changes caused by the addition of chloride, though in the present study they are observed to be small. The authors also connected such delocalization to the dynamic rotation of the organic MA$^+$ cation, and speculate that the lower PLQE observed in all-inorganic perovskites, such as CsPbI$_3$, might be due to the lack of this rotating cation.[187] However, at least one study suggests that, while the local dielectric environment changes due to the cation rotation, the charge carrier is unaffected,[188] suggesting that there is more work needed to bring experiment and theory in agreement on the impact of defects on optoelectronic properties.

Non-native defects have also been incorporated into perovskite thin films.[189] Many serve the purpose of modifying the microstructure through tuning the crystal nucleation density and growth kinetics and are thus discussed in section 5 for their function as additives.[190,191] However, some do have a distinct impact on the optoelectronic properties of the films by means of altering the electronic structure, the crystal structure, and the carrier lifetime recombination dynamics.[192] It should be noted that the distinction between the full substitutions/alloys discussed in section 3.1 and those discussed here is not perfectly clear. However, we consider replacement limited to small percentages to be 'non-native defect introduction' and distinct from those discussed previously.

One avenue has been to partially substitute the lead at the B-site with other transition metals. For example, it was found that strontium (Sr) insertion into the perovskite lattice was beneficial to solar cell performance up to a replacement value of 2% of the lead by means of improving the carrier lifetime.[193] This is supported by an improvement in the fill-factor of the devices up to a remarkable 85%; however, the $V_{OC}$ simultaneously decreases with increasing $Sr^{2+}$ due to a decrease in the built-in potential of the devices. Klug *et al.* performed a survey study on B-site substitution (i.e. partial replacement of the lead) with interesting results.[194] Using the one-step lead acetate recipe (Pb(OAc)$_2$) as a base,[86] they partially substituted the $Pb^{2+}$ with other divalent B' cations (where B' = $Co^{2+}$, $Cu^{2+}$, $Fe^{2+}$, $Mg^{2+}$, $Mn^{2+}$, $Ni^{2+}$, $Sn^{2+}$, $Sr^{2+}$, and $Zn^{2+}$) such that the final composition of the perovskite was MA(Pb:B')I$_3$. The Pb:B' ratio was varied between 1.6% and 25% replacement of the lead. They found that the perovskite generally tolerates replacement of up to 6% of the lead except for $Fe^{2+}$, where any amount introduced trap states and diminished photovoltaic performance. $Sn^{2+}$ and $Co^{2+}$ had the largest benefit at low-level substitution via enhancement of the $V_{OC}$. The impact of these substitutions on the (intensity normalized) photoluminescence, shown in the supplementary files for the paper, was varied, with $Ni^{2+}$, $Sr^{2+}$, $Fe^{2+}$, and $Mg^{2+}$ having a strongly blue-shifted peak position at high replacement levels, and the others remaining close to the MAPbI$_3$ control value of 774 nm for all impurity levels. Related to this shifting, the addition of impurities was also found to impact the crystal structure of the perovskite.

The replacement or addition of $Bi^{3+}$ to the perovskite lattice has received varied experimental and theoretical attention, with intriguing results. Its inclusion in the MAPbI$_3$ lattice has been shown to have a strong impact on its photophysical properties, perhaps most interestingly for its ability to act as a luminescent center in the near-infrared (NIR) regime, which is necessary for applications in telecommunications. Here, Zhou *et al.* found that for increased doping levels of 0.005-0.25%, the photoluminescence intensity ($\lambda_{exc}$ = 517 nm) gradually decreased at the 780 nm peak but increased at 1140 nm (**Figure 10**a).[195] Also shown in Figure 10a, the TRPL reveals increasingly fast and dominant initial decay, indicating a change in the deexcitation channels of the carriers. The authors suggest that this supports the mechanism of photoexcitation via energy transfer from the MAPbI$_3$ semiconductor matrix to the bismuth-related active center, depicted schematically in Figure 10b. As a proof of concept of their results, they fabricated LEDs of the Bi-doped perovskite, which showed EL in the range of 1100-1600 nm at 84 K. On another positive note, Abdelhady *et al.* found that MAPbBr$_3$ single crystals doped with Bismuth exhibit an increase in conductivity as a function of doping level, thus also indicating that B-site substitution can engender tunable functionality.[196] However, other studies indicate that this tunability does not extend to optimal device operation. Yavari *et al.* examined impact of Bi$^{3+}$ inclusion from a different angle, studying a range of impurity levels to deduce how tolerant their triple-cation, mixed-halide films are to this particular defect.[197] They found that any amount of Bismuth decreases the open-circuit voltage of solar cells, reduces both EL and PL, and increases the Urbach energy of the film. DFT calculations suggest that this is because the Bismuth substitution at the lead site forms both a deep and a shallow trap state, thus simultaneously being a center for non-radiative recombination, reducing the charge transport in the film, and causing an increase in the energetic disorder via broadening of the tail states (Figure 10c). Single crystal studies on this defect in MAPbBr$_3$ come to a similar conclusion; here, researchers found that despite the apparent shift of the absorption onset and the color change of single crystals (shown in Figure 10d), the band edge does not shift with 1-10% Bismuth inclusion.[198] Rather, the impurity causes an increase in the density of states near the band edge, more than doubling the Urbach energy and lowering both the PL intensity and lifetime.

Viewed altogether, the results of these studies first indicate that the theoretical and experimental picture of defect physics in perovskites and its impact on luminescence is neither complete nor unified. We expect further studies which attempt to bring the two into agreement will be highly beneficial to the field. Secondly, these results suggest that careful control over impurities in the perovskite cells may be beneficial to devices in the form of so-called "defect-engineering", promoting luminescence in a particular regime, such as the NIR, or enhancing luminescence. Further efforts to control the occurrence of specific defects and mitigate non-beneficial behavior via processing and post-processing techniques are an important step in this fabrication effort and are discussed further in section 5. It is important to note, however, that this represents only an initial snapshot of the impact of native and non-native defects on the emissive behavior of a film. As we discuss in sections 4 and 6, defects interact strongly with an electric field, migrating over the length of an entire film, and additionally interact with the environment. Both effects cause a strong change in the photophysics of the film over time. In all cases, there is a further compositional variable, which has an impact on the photoluminescence: the microstructure of the perovskite film. Hence, this is the topic of discussion for the next section.

### 3.4. The Impact of Microstructure

Most commonly, photoluminescence measurements on perovskite samples represent an average acquired over different regions/domains of the (most often) polycrystalline perovskite films. Nevertheless, perovskite films are intrinsically non-uniform and exhibit domains of different sizes, whose properties depend on the exact material composition and the processing technique parameters.[199,200] This heterogeneity is observable in the photoluminescence properties of the film.[59,201–203] Characterization of the photoluminescence properties of perovskite films on the microscale is of particular importance for the development of a deeper understanding of the local physical and chemical processes taking place in this complex class of materials. Photoluminescence mapping allows for the spatial resolution of the photoluminescence properties of individual domains in the perovskite film and also for mapping the local differences in PL intensity, lifetime, or emission wavelength, depending on layout of the setup and the chosen detector. For wide-field mapping of PL intensities, the film is placed in the focal plane of an objective through which it is excited with short wavelength laser light.[204,205] The same objective is then used to collect the long wavelength photoluminescence, which is separated by a dichroic mirror and long-pass filters from the excitation beam and fed onto a CCD camera. Using a confocal microscope with either a laser scanning or a piezo-driven stage dramatically increases the lateral resolution, up to the diffraction limit.[206,207] Using a photon counting detector and a pulsed excitation laser allows for the direct determination of spatially- and time-resolved photoluminescence.

Combining PL mapping with visualization methods of the local microstructure such as atomic force microscopy (AFM), scanning electron microscopy (SEM) or micro X-ray diffraction (μ-XRD) leads to an enhanced understanding of the local properties and processes in perovskite films that determine their photoluminescence and accompanying spatial heterogeneity.[199,207] It is worth mentioning that when interpreting differences in PL, or analyzing the heterogeneity of PL maps, special attention must be paid to the differences in film roughness and thickness, especially by comparing films prepared by different fabrication methods. A higher surface roughness, or large variations in film thickness, can lead to significant differences in light outcoupling and self-absorption, changing the spatially observed PL intensity, peak position, and shape. These effects further complicate a comprehensive and complete analysis and comparison of PL properties between different deposition techniques and research labs, as well as between different perovskite materials. In the following section we will outline some of the most noteworthy results obtained using PL microscopy.

The spatial heterogeneity in PL maps is characterized through substantial differences in the microscale emission – i.e. mapping areas of high and low PL intensity. The pioneering studies of de Quilettes *et al.* in 2015 demonstrated that the PL intensity within a homogenous compact film of a $MAPbI_{3-x}CL_x$ varies by up to 30% between neighboring grains.[59] By correlating time-resolved measurements at different intensities of bright and dark areas, the authors demonstrated that dark grains exhibit a higher trap density and more nonradiative pathways. In particular, grain boundaries, observed using SEM, exhibit a 65% lower PL intensity and significantly shorter PL lifetimes (**Figure 11**a-c). These findings indicate that grain boundaries are defect-rich regions and centers of nonradiative recombination and not as benign as earlier results suggested.[65] Treatment of the films with pyridine brightens the overall PL but enhances the initially dark areas significantly more than the areas that were bright from the onset, suggesting that pyridine mediates the local defect and trap concentration (see section 5.2), particularly at the grain boundaries. To relate the observed heterogeneity in PL with device performance, Eperon *et al.*[208] combined spatially resolved photovoltage and photocurrent measurements on MAPbI$_3$ films with PL mapping. An anticorrelation between PL intensity and both photocurrent and photovoltage, and hence efficiency, was found, calling into question the commonly accepted belief that a high $V_{OC}$ can only be achieved in highly luminescent samples. Using different contact materials in the same study, the authors demonstrated that the device heterogeneity stems from the perovskite material, possibly from variations of the surface composition, and not from the contact material.

It has been a topic of heavy debate whether the overall contrast in PL maps is linked to nonradiative recombination in the perovskite material or to diffusion outside the confocal laser spot. Jin and co-workers demonstrated that photoexcited carriers generated through a diffraction-limited spot diffuse outside the studied location, complicating PL analysis.[209] Zhu and co-workers showed by kinetic modelling that the grain boundaries play only a minor role in determining the nonradiative recombination and exhibit similar lifetimes compared to the inner grain area when diffusion after excitation is considered.[210] Final proof that the heterogeneity of the PL map is dominated by nonradiative recombination and not by diffusion outside the excited laser spot was given by Ginger and co-workers by comparing confocal PL mapping of different intensities with wide-field imaging (Figure 11d-e).[211] Importantly, the same study also confirmed that the overall observable contrast and heterogeneity within PL maps is a function of excitation intensity. Both mapping techniques provide similar insights into the PL properties as long as the excitation powers are low (typically 1-5suns) and comparable with excitation densities required for solar cell operation. The extent to which grain boundaries remain detrimental to high performance in perovskite devices and limit the $V_{OC}$ through nonradiative recombination remains a controversially discussed question, subject to an ongoing debate.[212,213]

Several suggestions have been made to explain the intrinsic heterogeneity of perovskite PL maps. Apart from the apparent lower photoluminescence of the boundaries between the domains, the domains exhibit a large spread in intensity among themselves, which empirically does not correlate with obvious parameters such as size or height. Since perovskite films consist of crystalline domains of different orientation, the differences in PL intensity could stem from differently oriented crystallites. To investigate this possibility, studies performed on single crystalline perovskites can help to correlate the photoluminescence properties with crystal orientation. It has been shown, for example, that $MAPbI_3$ exhibits facet-dependent optoelectronic properties and efficiencies in solar cells.[214–217] Xu *et al.* showed that for sequentially grown $MAPbI_3$ crystallites the photoluminescence properties are different depending on the nature of the crystal facet and the presence of structural defects. Completely smooth facets show a PL maximum at 775 nm while step-like crystal facets exhibit emission that is blue shifted by approximately 5 nm.[218] Facet-dependent photoluminescence of $MAPbI_3$ has been investigated by Kim *et al.* by exciting the (100) and (112) crystal facets of large $MAPbI_3$ single crystals[184]. As described in more detail in section 3.3, the authors observed differences in both the PL intensity and peak position between the different crystal facets. These differences were ascribed to facet-dependent defect types, which leads to variations in nonradiative recombination and marginal changes in the band gap. These findings stand, however, in contradiction to the observations made by Ehrler and co-workers who studied the photoluminescence properties of large, highly oriented $MAPbI_3$ crystalline domains grown by flash infrared annealing (FIRA).[219] The authors did not observe changes in PL intensity or spectral position between highly ordered domains with different crystallographic orientation.

The differences in the microscopic domain size might intuitively be the most obvious explanation for the observed heterogeneities in the PL of perovskite films, such as differences in PL peak position or radiative lifetime. Early studies by Pettrozza et al.[220] indeed reveal that differently sized crystallites of MAPbI$_3$ exhibit very different PL properties. Small crystallites grown in an oxide scaffolding of just tens of nanometers in size show a larger band gap, therefore a significantly blue-shifted PL. The PL decay rates also vary with crystal size, with smaller grains exhibiting a shortened lifetime, while larger crystallites on top of the oxide scaffolding exhibit a significantly longer lifetime in direct comparison. The same relationship has been observed for MAPbI$_3$ crystallites deposited on glass ranging from below 250 nm to >2 µm in size with PL lifetimes of 2ns up to 100ns.[221,222] Excitation intensity dependent PL measurements on these films revealed that the differently sized crystallites exhibit different bimolecular intrinsic radiative recombination coefficients and consequently a different PLQE. The authors attribute the changes in the optical bandgap, and hence the differences in PL, to differences in lattice strain between the small and large crystallites. Longer radiative lifetimes in larger grains were also described by Rumbles et al.,[223] who studied the microwave photoconductance of MAPbI$_3$ films with varying grain sizes. However, the spectral position of the PL peak does not change with grain size in their study. Interestingly, besides recombination of charges being essential in both of the decay processes, the PL lifetime of the differently sized grains does not correlate with their transient photoconductivity. The latter is significantly slower in small grains, opposing the trend observed in the PL measurements. The relationship between PL and grain size is also part of the aforementioned study of Ehrler et al. in 2019.[219] Here, the photoluminescence peak position of large (tens of micrometers in size) FIRA-grown MAPbI$_3$ grains is fully comparable to smaller grains (hundreds of nanometers) in a film fabricated by the antisolvent method, after the PL data was corrected for self-absorption effects. Despite the differences in the overall size, the PL properties of MAPbI$_3$ did not change in their studies. Similarly, studies by Yan et al. did not observe a universal correlation between grain size and spectral PL peak position of MAPbI$_3$.[224] Therefore it remains unclear to which extent the crystallite size of MAPbI$_3$ impacts its PL properties.

Very recent studies by Jariwala *et al.* combined extremely sensitive electron backscatter diffraction (EBSD) on MAPbI$_3$ perovskite films with high resolution PL mapping, revealing that the heterogeneity present in photoluminescence maps is directly related to the inter and inner grain misorientation.[206] The study emphasizes that structures visualized by SEM, and commonly referred to as 'grains' (even within this article), are not grains in a crystallographic sense. SEM visualizes the morphology of a sample but does not provide any information about diffraction or crystallographic orientation. Only through the development of a sensitive detector could beam damage in the perovskite film during EBSD characterization be avoided, and meaningful experiments could be conducted using this method.[225] By employing the EBSD technique, Ginger and co-workers[206] recently revealed that large grains observed in SEM typically consist of multiple sub-grains with individual orientations.[226,227]

The inverse pole figure (IPF) map obtained by EBSD reveals a large heterogeneity in crystallographic orientation within the perovskite film (**Figure 12a**). By defining a threshold that determines which data points are grouped as grains, a grain boundary network can be obtained (Figure 12b). These EBSD results unveil grain boundaries that were not previously visualized by SEM.[206] Furthermore, not only does the orientation vary greatly from grain to grain, a distribution of orientations is present also within individual grains. Defining the deviation from the average grain orientation as the grain orientation spread (GOS), the spatially resolved local structural and strain heterogeneity can be visualized (Figure 12c). Superimposing a confocal PL map with the grain boundary network obtained by EBSD, and analyzing the correlation between individual each grain's GOS with its PL intensity, reveals an anticorrelation between strain and PL intensity (Figure 12d). These findings are in line with the studies by Jones *et al.* who used μ-XRD to probe the microscopic lattice strain and correlate it with PL heterogeneity.[227] The origin of the lattice strain is considered to lie in the fabrication process of the perovskite material, which requires annealing temperatures of 100 °C. Cooling down back to room temperature causes a phase transition from the cubic to the tetragonal phase, creating strain between neighbors which transitioned at different times.[159] These results highlight that many factors contribute to the observed heterogeneity in the photoluminescence properties of perovskite materials and significantly more research is required to identify the causes of these variations and develop mitigation strategies.

Beyond visualization of the intrinsic heterogeneity of perovskite films, PL microscopy can also be used for the study of local and/or temporal effects in perovskite films. Light soaking has been shown to have a tremendous influence on the photoluminescence and performance of perovskite films.[58,228] Particularly, the light-induced compositional changes and redistribution of ions by light were extensively studied by PL mapping[229–231] and will be discussed in more detail in section 6, however we now turn our attention to the monitoring of the electric field-induced ion migration performed by PL mapping.

## 4. Using PL *in situ* as a Probe for Ion Migration

One of the more intriguing applications of PL in perovskites is in the measurement and quantification of ionic motion. In this section we will provide a brief history of the development of this technique, as well as a summary of the most prominent results, as they pertain to perovskite devices.

### 4.1. Background

Early research into the application of perovskite solar cells discovered significant J-V curve hysteresis,[27] which was originally attributed to either a ferroelectric effect[150] or to mobile ions drifting through the layer in response to the applied voltage. However, later efforts that examined the effect in more detail found that the low-frequency hysteretic charge density was too large to be caused by ferroelectricity,[232] and combined with the report of an *in operando* switchable photocurrent direction in 2015,[233] these results pointed towards mobile ion species as the culprit.

This ion migration is speculated to be responsible for many of the unique and interesting properties of perovskite materials,[234] for example the light induced self-poling[235] and phase-separation[236] effects, or for its potential application in energy storage devices.[237] However, the low formation energies of many defects, especially X-site point defects,[11,29,238,239] gives rise to a significant amount of ionic motion, enough to cause not only the aforementioned hysteresis, through screening of the built-in potential, but also accelerate device degradation.[43,240–244] Therefore, an accurate measurement of the ionic motion in perovskite materials is needed to further our understanding of this complex system.

PL measurement techniques, especially time-resolved PL, are an excellent candidate to aid in this task, as the motion of ions necessarily changes the type and nature of the defects within the crystal lattice. As discussed previously, if these defects are located deep enough within the band gap, they can act as nonradiative recombination centers and will therefore modulate the PL signal. This allows for researchers to use PL microscopy as a pseudo-direct visualization of the presence, absence, or movement of ionic species.[231,245]

### 4.2. History

Even early in the research into perovskite materials, time-resolved PL measurements began to indicate the mobile nature of some ions. As early as 2014, Sanchez and coworkers[246] found that the PL intensity of MAPbI$_3$ films increases both when changing the deposition from single to two-step methods, as well as by replacing the MA$^+$ cation with FA$^+$. More interestingly, when examining the transient and steady state evolution, they observed unique behavior for all three scenarios with slow (second to millisecond) time scales. While they could not assign a mechanism to this phenomenon, they speculated that it could be related to the J-V hysteresis issue. One year later in 2015, Qiu *et al.*[247] observed spontaneous PL switching in electric field-dependent PL transient measurements of MAPbBr$_3$ nanoplatelets on 1-10 s time scales, which are similar to values reported for ion drift in perovskite thin films. That same year, Leijtens *et al.*[248] found that the application of an external electric field in laterally interdigitated devices enhances the PL signal by a reduction in the monomolecular trapping rate, which they proposed originates from a removal of excess ion interstitials – "leftover" precursor molecules from an imbalanced stoichiometry. Further evidence that ions are responsible for these variations in PL properties was provided by de Quilettes and coworkers,[59] also in 2015, who compared the PL intensity of specific grains against the Cl:I ratio in mixed-halide MAPbI$_{3-x}$Cl$_x$ films obtained by SEM and energy-dispersive X-ray spectroscopy (EDX). They found a positive correlation between higher amounts of Cl$^-$ and PL intensity, further strengthening the link between the local PL properties and ionic composition.

### 4.3. Spatial and Time-resolved Mapping

Subsequently, similar publications from several groups, including work by Chen,[230] Deng[204] and Li[205] *et al.*, attempted to quantify the type and mobility of the as-yet unknown mobile ion species in MAPbI$_3$ by employing time resolved, spatially mapped, electric field dependent PL microscopy. Here, in a similar device architecture to that used by Leijtens, an external electric field is applied to a perovskite thin film laterally by biased metal electrodes, as shown in **Figure 13**.

As the external bias is applied, mobile ions within the perovskite film drift towards the appropriate electrode via coulombic forces, and the resulting point defects (interstitial or vacancy) cause local PL quenching. By measuring the PL response for each point across the entire channel, the researchers were able to observe the movement of these ions *in situ*. These groups found that upon biasing, a quenched region forms at the anode (positively charged electrode) and progresses toward the cathode (negatively charged) over time. The speed at which this quenched region expanded was found to be linearly proportional to the strength of the electric field, consistent with a simple model of mobile ions.[249] This quenching was found to be reversible for certain atmospheres and field strengths.[204]

Interestingly, concurrent to these reports other groups have observed the opposite phenomenon: that under bias, a quenched region begins at the *cathode* and proceeds towards the anode.[250–252] This inconsistency could be caused a different initial doping state as suggested by Hüttner and coworkers,[253] as the surface of perovskite films have been shown to be extremely sensitive to precise precursor stoichiometries.[41,254] Alternatively, differing fabrication conditions could lead to significantly higher $MA^+$ diffusion coefficients, as Senocrate[255] and Shao[256] suggest that, under certain conditions, $MA^+$ ions could diffuse through grain boundaries more rapidly than through the bulk. This ion migration would naturally progress in the opposite direction due to the opposite charge on the $MA^+$ cations.

Regardless, the results of the majority of studies, those mentioned above and others,[245,253,257] concluded that the dominant mobile ion species in perovskite thin films is most likely the halide, specifically vacancy type defects, which are able drift throughout the film in response to an applied external field and enhance the local nonradiative recombination rate, causing the observed PL quenching.

In order to extract the mobilities for this motion, the position of the quenching front as a function of time and field strength can be used. Complicating this measurement is the difficulty in assigning a definite position to the quenching front, as often it is rough and undefined. In a recent publication,[258] our group exploited the grain microstructure uniformity of zone-cast perovskite ribbons to make extremely precise measurements of the quenching front progression, as seen in **Figure 14**.

By fitting the position of the quenching front with respect to time, and using the applied external bias of 0.9 V µm$^{-1}$, we were able to extract a mobility value 3.8 x 10$^{-9}$ cm$^2$ V$^{-1}$ s$^{-1}$, leading to a diffusion coefficient of 9.8 x 10$^{-11}$ cm$^2$ s$^{-1}$, in good agreement with the reported values.[204,259] Up until now, we have covered the use of PL in elucidating the effects composition, microstructure, fabrication, and defects have on perovskite films. In the next section, we move from observation to active strategies designed to enhance performance.

## 5. Advancing Film and Device Fabrication Techniques Using PL Spectroscopy

As we have discussed in the preceding sections, PL is closely related to the defect content and microstructure of the perovskite film. Since these qualities are a direct result of the processing method, measuring the photoluminescence as a function of the fabrication recipe is a powerful way to understand the impact of any modifications or additions. Much of this is accomplished with the goal of defect management, aiming to leverage the positive traits and mitigate the negative. Recipe modifications fall into a few general categories, many of which have already been reviewed extensively.[31,190,260] Here, we give an overview of those in which the PL has played an especially important role, either because it was a tool for increased understanding or because its intensity was maximized through a specific fabrication procedure. We first discuss recipe changes which occur during the processing of the initial film via additives to the precursor solution or changes to specific parameters, followed by a discussion of post-processing through surface passivation techniques. As the non-radiative losses due to imperfections in the material are minimized, further improvements in the $V_{OC}$ will be accomplished via improved photon management strategies. These are already proving promising for perovskites, and are the final topic of discussion for this section.[33] Insights into the processing window are especially critical in the development of scalable methods for perovskites.[261,262] This remains a challenge for perovskites, which show a performance drop as the device size and active area of a film are both increased.

## 5.1. Using PL to Manage Defects Incurred During Processing

### 5.1.1. Recipe Optimization

The solution-processing routes to a functional perovskite film are surprisingly diverse. For spin-coating, they can be two-step processes, where one precursor is deposited as a film and the other intercalates into it to form the perovskite intermediate, or one-step processes, where all precursors are deposited onto the substrate at once.[31,84,260,263] In each case, there are several parameters which must come together in order to form a film with a given microstructure and defect content. For example, a given recipe might call for an antisolvent to form the intermediate perovskite phase. Here, the chemical nature of the solvent used (e.g. boiling point, polarity) and the point at which it is deposited during the recipe[264] both impact the crystallization kinetics of the film. Other parameters of interest are the annealing time and temperature, the impact of the fabrication environment, the solvent for the precursor solution,[265] and more. Examining these processes as they happen – i.e. *in situ* – has been a successful approach for optimization and understanding of organic photovoltaics[266,267] as it reduces the guess-and-check nature that might come with analysis after the fact. Such analysis can also yield illuminating results for perovskites.

Many perovskite recipes require a thermal annealing step after spin-coating to convert the intermediate phase to the photoactive phase,[260] which can vary in both duration and temperature. To better connect annealing time with the final device performance, van Franeker *et al.* examined the photoluminescence of MAPbI$_3$ films on the hole transport layer poly(3,4-ethylenedioxythiophene) polystyrene sulfonate (PEDOT:PSS) as they were forming on a hotplate.[32] As shown in **Figure 15**a,b, they monitored the intensity of the PL at 770 nm (in the vicinity of the peak PL for MAPbI$_3$ films) for different hot-plate temperatures, finding that the PL quenches more quickly at higher annealing temperatures. They observed that the maximum PCE of finished devices corresponds to the annealing time when the PL first reaches its base value (where it has just been quenched, shown in Figure 15c), reasoning that this occurs at the point when the film quality is optimal for charge transport. At this point, charges excited at the top surface of the film have a high enough diffusion length to reach the hole extraction layer, where they non-radiatively recombine.

Photoluminescence has also been used to more directly understand the kinetics of the intermediate phase formation and its relationship to the final device properties. Li *et al.*[268] examined the impact of humidity on the formation of MAPbI$_3$. After depositing the precursor solution by spin-coating, the samples were immediately placed in a humid (0%, 35%, 63%, 83%) N$_2$ atmosphere and kept at room temperature. Monitoring the samples' PL response over time revealed an initial complex, multiband signal that approaches the reference spectra. Increasing the level of humidity hastens this transformation until 83% humidity, wherein the device rapidly degrades into a transparent photoinactive phase. Using a fitting model consisting of multiple Voigt profiles, the authors found that the initial multiband signal is made up of five distinct phases. It was speculated that these are caused by stable, low-dimensional water-perovskite complexes, as these signals are all blue shifted with respect to the reference, possibly due to quantum confinement effects.

While these two examples illustrate how monitoring of PL during perovskite film formation can be a valuable tool to understand and optimize its properties, the complexity of implementing a PL characterization setup directly adjacent the fabrication area prevents more detailed studies of this kind. Consequently, most PL studies are performed after the film has already been formed, with its properties modified either during or after processing. In the following we survey some of the most striking examples of such approaches.

*5.1.2. Additive Chemistry*

Even though perovskites are reported to be defect tolerant, it still seems counterintuitive that adding an extra chemical to the precursor solution – which may or may not remain in the final film – would improve PL and device performance by means of defect management. Nonetheless, several reports have indicated that this is indeed the case. In fact, there are so many reports of additives that it might be more productive to ask what materials have not been mixed with perovskites. This topic has been reviewed extensively by Bo Li *et al*.[269] and Taotao Li *et al*.[191] where they have discussed polymers, solvents, fullerenes, nanoparticles, metal or organic halide salts, inorganic acids, and more, with a range of benefits including enhanced crystallization kinetics for more uniform grains and defect passivation. As this list of potential additives is too exhaustive to be covered here, we instead will present a series of experiments which utilized PL measurement techniques as a powerful tool to understand the additives' effect on the perovskite photophysics.

One common strategy used to enhance the performance of perovskite devices is to reduce the amount of residual metallic lead within the film, which acts as a nonradiative recombination center.[270] Zhang *et al*. found that the addition of hypophosphorous acid (HPA) [185] to the one-step lead-acetate recipe to fabricate $MAPbI_3$ films[86] reduces the amount of metallic lead in the film detectable via XPS. Along with this, the authors found that the addition of HPA results in an increase of the steady state PL and TRPL decay constant by an order of magnitude. Cho *et al.* used a 5% molar excess of MABr[271] in the precursor solutions of $MAPbBr_3$ for light emitting applications, and boosted the current efficiency of their LEDs by two orders of magnitude compared to an exactly stoichiometric precursor solution. Using the additive 2,2′,2″-(1,3,5-benzinetriyl)-tris (1-phenyl-1-H-benzimidazole) (TPBI) in conjunction with a chloroform antisolvent treatment, they reduced the average crystal grain size via nanocrystal pinning, resulting in superior outcoupling and an even higher current efficiency (up to 8.53% EQE). In a similar vein, Xie *et al.*,[272] inspired by the long diffusion lengths observed in MAPbI$_{3-x}$Cl$_x$,[9] instead used MACl as an additive to their MAPbI$_3$ perovskite prepared by an MAI:PbI$_2$ recipe, and characterized the resulting films by a range of PL techniques. Examining the TRPL decay for films on glass and both electron and hole transporting layers, the researchers found longer PL decay coefficients on glass, but shorter on transport layers. Increasing the excitation fluence up to 600 μJ cm$^{-1}$, amplified spontaneous emission (ASE) from the films was also measured. Both a lower ASE threshold power density and narrower ASE emission peak was found for increasing MACl amounts. These results are all consistent with a reduction in nonradiative recombination pathways, which was verified via significantly improved PV performance. In a corresponding experiment, Dänekamp and coworkers modified the ratio of deposition speeds of MAI and PbI$_2$ using a vacuum evaporation technique. They found that after exposure to ambient air, the PL lifetime and intensity is massively enhanced when the relative rate of MAI:PbI$_2$ deposition is 3:1 versus 1:1. Using this information, LEDs fabricated from the 3:1 deposition rate ratio possessed an EQE of 2%, 40 times higher than the control devices[273].

Wu and coworkers[274] found that additions of methylammonium acetate (MAOAc) and thio-semicarbazide (TSC) to the MAI:PbI$_2$ precursor solution dramatically enhanced the coverage and grain size of the resulting films. Time resolved and steady state PL revealed enhanced electronic properties in the form of longer decay constants and a slight blue shift in the PL peak position, suggesting the removal of in gap trap states which could red shift emission.[275] To make an estimate of the actual change in the defect density, the authors employed temporal-integrated PL (TIPL), which relates the charge carrier density within the film to the time-integrated PL intensity after an excitation laser pulse. At low fluences, Auger recombination can be neglected, and so only radiative and trap-assisted recombination are possible. The initial charge carrier density $n_c$ can be estimated via multiplying the laser fluence by the film's absorption coefficient, and then the defect density can be found by the fitting formula (**Equation 16**)

$$n_c = \sum_i n_{TP}^i \left(1 - e^{\left(-a_i \tau_0 I_{PL}/k\right)}\right) + I_{PL}/k \qquad (16)$$

where $n_{TP}^i$ is the density of the *i*th trap, $a_i$ is the product of trapping cross section and charge carrier velocity, $\tau_0$ is the PL decay lifetime, and k is a fitting constant. Using this technique, two different trap types were needed to fit the data, which the authors ascribed to slow trapping "surface" and fast trapping "bulk" defects. As shown in **Figure 16**, the TSC additive treatment was shown to reduce both densities by an order of magnitude, indicating substantially improved electronic quality.

Another popular class of additives are metal halide salts, for example the Lewis acid-base coordination complex aluminum acetylacetonate Al(acac)$_3$. Wang *et al.*[276] found that "doping" MAPbI$_3$ films with the compound improved the performance by specifically passivating defect sites at the grain boundaries and reducing the film microstrain. They verified this in part by TRPL and PLQE in concert to make estimates of the radiative ($k_1$) and nonradiative ($k_2$) decay rates, finding that while undoped and doped films exhibited similar $k_1$, the addition of 0.15 mol% Al(acac)$_3$ to the precursor solution reduced $k_2$ by a factor of two, confirming the proposed mechanism for improved performance. Also heavily employing PLQE were Stoddard and coworkers,[44] who used a bespoke measurement setup to simultaneously measure absolute PL in conjunction with a four-point probe system to measure the photoconductivity. Using the known flux from a calibrated LED excitation source, PLQE and mean diffusion length L$_D$ could then be calculated. Using this setup to measure the oxygen degradation of MAPbI$_3$ and the halide segregation in MAPb(I$_{0.66}$+Br$_{0.34}$)$_3$ films, they uncovered a seemingly counterintuitive phenomenon whereby PLQE could be increasing but L$_D$ decreasing. They attribute this to the formation of local regions of increased charge carrier density: either by degradation or halide segregation, areas of higher band gap material (PbI$_2$ or Br$^-$-rich perovskite) generate photoexcited carriers, which then diffuse to the lower band gap perovskite and become trapped. These regions exhibit higher radiative recombination, but as the charges are stuck in an energetic well, the $L_D$ is concurrently reduced. The authors emphasize that a measurement of PLQE alone could lead to an incorrect conclusion regarding what is actually happening within the film. The authors then go on to apply this measurement technique to analyze the addition of the Lewis bases triethylamine (TEA) and trioctylphosphine oxide (TOPO) to the fabrication of mixed cation mixed halide lead perovskite $FA_{0.83}Cs_{0.17}Pb(I_{0.66}+Br_{0.34})_3$ films used in PV devices. Correlating their findings regarding PLQE and $L_D$ to device performance, they were able to balance the competing effects of increasing PLQE and decreasing $L_D$ at higher additive concentrations to concentrations to improve both $V_{OC}$ and short circuit current $J_{SC}$. Another noteworthy example is the addition of potassium iodide into triple cation-based perovskites reported by Stranks and coworkers.[277] By optimizing the fraction of potassium in their films, the authors demonstrated that both the external and internal PLQE can be drastically optimized, reaching values of 66% and 95% respectively. This increase was accompanied by the removal of the fast PL decay component upon passivation, which is associated with nonradiative decay pathways. Strikingly, the authors demonstrated that these excellent PL properties are highly stable, with continuous illumination leading to nearly no changes in neither PLQE nor PL peak position and were maintained even upon interfacing with charge extracting layers.

**5.2. Using PL to Mitigate Defects After Processing via Surface Passivation**

No matter how well-grown the bulk crystal, film, or domain, it must terminate at a surface or grain boundary, where there will naturally be a high density of defects to act as non-radiative recombination centers. Surface passivation has been an effective strategy for defect management in all solar harvesting materials, promoting its use to mitigate losses in perovskites. Here, however, the efficacy of the strategy depends on the chemical nature of the defect. Perovskites will terminate in under-coordinated ions; but these can be positively charged $Pb^{2+}$, which requires charge-donating molecules for passivation, or negatively charged halide vacancies, which require electron-acceptors for passivation. As a result, several materials, ranging from Lewis acids to bases, as well as organic halide salts, have proven to work well, enhancing both the PLQE and the PL lifetime of the bulk material and the $V_{OC}$ in a device.

Noel *et al*. treated the surfaces of $MAPbI_{3-x}Cl_x$ films with the Lewis bases thiophene and pyridine, finding that for both treatments the external PLQE ($\lambda_{exc}$ = 532 nm, variable fluences, encapsulated with PMMA) is improved over the control film, no matter what the fluence.[278] They recorded the PLQE at 1 sun to be nearly 25% for the pyridine-treated samples, 20% for the thiophene-treated samples, and 15% for the control samples. TRPL measurements demonstrated that the surface treatments enhance the PL lifetime to approximately 2000 ns in both cases, with the thiophene treatment being slightly higher, supporting the argument that the density of non-radiative decay centers is reduced. The authors ascribe the mild discrepancy between the PLQE and TRPL results by suggesting that the radiative decay rate is also impacted by the treatment. When incorporated into devices, they were shown to enhance both the $J_{SC}$ and $V_{OC}$. Similarly, Braly *et al*.[37] spin coated the Lewis base TOPO onto already fabricated $MAPbI_3$ films and observed the PLQE to rise to 42%, from the 1.2% of a control device. Using a technique to calculate the internal PLQE (i.e. eq. X), which accounts for photons generated within a film but unable to escape due to poor outcoupling, by varying the back-surface parasitic absorption (**Figure 17**), the passivated films were found to have an internal PLQE exceeding 90% for a 1 sun excitation intensity. Furthermore, the authors reported that the quasi-Fermi level splitting, and thus the inferred $V_{OC}$, obtained by the TOPO passivated films reached over 97% of the theoretical radiative limit, the highest reported at the time of publication, demonstrating the potential of surface passivation to maximize device performance.

Another common strategy is the use of amino-terminated passivating molecules,[279–281] which can anchor to $Pb^{2+}$ ions through coordination bonds, or to $I^-$ ions through hydrogen bonding, and thus passivate those sites. Xu *et al.* examined a variety of different amino-terminated alkyl chains and, using a combination of experimental data and first principles calculation, rationally design a molecule to passivate the surface of FAPbI$_3$ films. Using fluence-dependent PLQE, they observed only a small advantage in PLQE for the highest performing agent, 2,2′-(ethylenedioxy)diethylamine (EDEA), over the others. However, at low fluences, where nonradiative recombination dominates the charge carrier dynamics (due to its linear dependency on the charge carrier density), EDEA had an order of magnitude higher PLQE, indicating a substantial reduction in the nonradiative decay rate. When integrated into LEDs, the authors reported an ELQE increase over the control device from 7% to almost 20%. Phenethylammonium iodide (PEAI) was found to effectively enhance both the PL and EL of mixed organic lead triiodide FA$_{1-x}$MA$_x$PbI$_3$ films when spin-coated directly onto their surface.[282] When incorporated into solar cells, the device $V_{OC}$ was enhanced from 1.12 V to 1.18 V (which is 94.4% of the 1.25 V possible maximum) without significant change in the $J_{SC}$. The authors noted, using steady state PL on bare films, that the intensity was enhanced upon incorporation of the PEAI, however when the subsequent films were thermally annealed a reduction in the PL intensity was observed. Using this information in conjunction with X-ray diffraction (XRD) measurements, the authors concluded that during thermal annealing the PEAI was converted into the 2D perovskite phenethylammonium lead iodide (PEA)$_2$PbI$_4$, which was inferior at passivation than plain PEAI. When the passivated films were incorporated into LEDs and solar cells, the ELQE value predicted a $V_{OC}$ enhancement of 0.056 V (Equation 12, section 2), and the PV cells displayed an impressive certified PCE of 23.3%, indicating that the treatment greatly suppresses non-radiative recombination. They ascribe this to the effective passivation of iodide vacancies and under-coordinated surface lead species.

In perovskite QD systems, confinement and outcoupling effects play a significant role in the PL behavior, and high PLQE values approaching 90% were rapidly achieved.[283] In an attempt to increase that value to 100%, Koscher *et al.*[284] investigated passivation treatments using ammonium (NH$_4$SCN) and sodium (NaSCN) thiocyanate on CsPbBr$_3$ QDs. By simply mixing the thiocyanate powders into the QD colloidal solution, the authors found the steady state PL blue shifted slightly - consistent with a reduction of in gap states - and the PLQE approached 100%. More intriguing was the effect the treatment had on aged QDs, which already showed signs of degradation: both the steady state PL and PLQE were significantly enhanced, and remarkably both were restored to the values of their pristine, passivated counterparts. TRPL revealed multiple decay pathways present in the control dots, seen as multi-exponential transient decay curves. In contrast, the passivated samples (both aged and pristine) showed highly mono-exponential behavior, indicating a complete elimination of defect sites and that radiative recombination was the only remaining decay pathway. Further characterization via XPS and Fourier-transform infrared Spectroscopy (FTIR) suggested that the mechanism responsible was the removal of excess lead from the surface of the dots and replacement with a thiocyanate ligand in that location. Attempts by the authors to transfer the thiocyanate treatment to other halide containing perovskite QDs were unable to replicate their success, intimating that composition-specific treatments may be required.

## 6. The Impact of the Measurement Environment

Both the steady state and time-resolved photoluminescence of materials can be heavily influenced the chosen measurement conditions. For example, it is well established that tracking the excitation power dependence of PL can be used to elucidate the photo-induced carrier recombination mechanisms in a range of semiconductors.[285] Similarly, it is well known that PL quantum yield is strongly influenced by temperature, showing a significant increase upon cooling in most luminescent materials.[286,287] The atmosphere in which the samples are measured may also play a role, as simultaneous exposure to air and light may trigger loss of photoluminescence via compositional changes or aging or alternatively cause an enhancement via suppression of trap states. In the following, we will discuss in detail the reported influences of measurement conditions on the photoluminescence properties of lead halide perovskites and identify key experimental parameters that should be systematically reported in literature in order to allow for meaningful interpretation of PL measurements and their reliable comparison among different research groups. We note that the role of temperature is excluded from discussion, since the vast majority of PL measurements are undertaken at room temperature.

**6.1. Excitation Parameters**

An integral part of any photoluminescence characterization is the photoexcitation of the material. The experimental parameters of this excitation play a critical role in determining the outcome of the PL measurements. Among these parameters the excitation power is perhaps the most recognized. With increasing excitation power, the efficiency of luminescence increases until it saturates at high charge carrier densities.[18,54] Detailed description about the nature of recombination at different excitation densities was already provided in Section 2.1, but in short, at low excitation densities at which traps are not entirely filled, recombination is predominately monomolecular in nature. Increasing the fluence beyond the point at which all traps are filled results in a change of the recombination to bimolecular. Finally, increasing the excitation fluence further will lead to an Auger-type, third order recombination. Detailed characterization of PL as a function of excitation power can be used to investigate both recombination mechanisms and trap densities in perovskites;[278] however, it is also important to report the excitation power in routine PL measurements. In particular, when using PL measurements to explain the device physics of solar cells, it is advisable to adjust the excitation power to match 1 Sun conditions.

Beyond excitation power, other parameters have also been shown to influence both the initial intensity and the temporal evolution of PL. Recently Motti *et al.* demonstrated a number of such influences.[288] For example, changing the repetition rate of the excitation source could lead to either a decrease or an increase of the PL. This change of PL behavior was even more pronounced when the geometry of excitation changed (**Figure 18**a). Depending on whether the sample was illuminated from one side of the sample (either perovskite or glass), or both sides of the sample, opposing trends of increase and decrease of PL were observed. The authors also demonstrated that that the evolution of PL under vacuum conditions also depends on the excitation energy, assigning the differences to the variation in the penetration depth of light of different excitation wavelengths. An earlier work by Quitsch *et al.* also highlights the importance of the choice excitation energy.[289] The authors demonstrated that either an enhancement or a decline in PL intensity can be observed for $MAPbI_3$ measured in air (Figure 18b) with a clear threshold at an excitation wavelength of 520 nm ($E$=2.38 eV). Considering that this threshold coincides with the bandgap of $PbI_2$, the authors conclude that this effect originates from the direct excitation of the residual $PbI_2$ in the perovskite layer.

## 6.2. Atmosphere

The remarkable sensitivity of the emission properties of lead halide perovskites to their environment was recognized early on. Despite numerous reports exploring the interaction of perovskites with their environment, much remains unknown about the exact mechanisms that govern these interactions. In this section, we will review the current understanding of the environmental influences on photoluminescence, focusing in particular on exposure to vacuum, $N_2$, $O_2$ and $H_2O$. We note that while it was shown that environmental conditions during fabrication might also have significant effects on the PL properties of perovskite materials,[290–292] herein we focus solely on the role of environmental conditions during PL characterization.

*6.2.1. Vacuum*

One of the most striking examples demonstrating the effect of vacuum exposure on the photoluminescence properties of lead halide perovskites was reported by Fang *et al*.[293] The authors investigated single crystals of MAPbI$_3$ perovskite and demonstrated that upon changing the measurement environment from the ambient to low vacuum conditions (~$10^{-4}$ mbar) the PL intensity is reduced by over two orders of magnitude. Concurrently, the PL lifetime is drastically shortened. Remarkably, subsequent exposure to the ambient fully reverses this behavior. The authors demonstrate that oxygen and water molecules physisorb on the crystal surface, effectively passivating the surface trap states and increasing the photoluminescence. A similar decrease in PL of MAPbBr$_3$ single crystals upon exposure to vacuum was also reported by Zhang *et al*., with the authors suggesting that oxygen and water molecules not only physisorb on the perovskite surface, but also penetrate into the crystal[294]. While Motti *et al*. reported a similar vacuum-induced quenching of PL on MAPbBr$_3$ and CsPbBr$_3$ thin films,[295] the authors further showed that a partial recovery of the PL in vacuum could be achieved by keeping the sample under dark conditions. This observation suggests that PL quenching in vacuum is not solely related to desorption of oxygen and water molecules from the perovskite sample, but also to its continuous photoexcitation. The authors suggest that photo-induced emissive deep trap states are responsible for the observed PL quenching under vacuum conditions, indicating an intrinsic photo-instability of perovskite materials.

Interestingly, despite these observations, several groups performed PL characterizations of perovskite materials under vacuum conditions showing both high intensity PL and relatively long lifetimes. For example, Liu *et al*. performed extensive temperature dependence studies of the steady state and time-resolved PL of MAPbI$_3$ and MAPbBr$_3$ thin films under vacuum (below $10^{-5}$ mbar) and did not report any loss of PL upon exposure to vacuum.[296] Moreover, G.

Grancini *et al.* observed long PL lifetimes in MAPbI$_3$ single crystals under vacuum conditions, which were reduced upon exposure to the ambient and fully recovered upon subsequent return to vacuum conditions.[297] Mosconi *et al.* reported a significant enhancement of PL in polycrystalline MAPbI$_3$ under illumination in vacuum and attributed this observation to light-induced annihilation of Frenkel defects.[298] These seemingly contradicting observations were recently reconciled by a pivotal experimental and theoretical work by Petrozza, De Angelis, and co-workers.[288] Remarkably, the authors showed that both a decrease and increase of PL can be observed under vacuum depending on the interplay between the defect formation and the healing processes that occur concurrently under illumination. It was convincingly demonstrated that varying experimental parameters such as excitation repetition rate, intensity and geometry may influence these competing processes, leading to opposing trends of PL under vacuum.

*6.2.2. Nitrogen*

One of the earliest studies on the effect of nitrogen on PL of MAPbI$_3$ thin films was reported by H. Míguez and coworkers in 2015.[299] Strikingly, the authors showed that exposure to N$_2$ under illumination can either enhance or reduce the PL intensity, depending on the history of the sample. Specifically, a pristine MAPbI$_3$ sample measured under N$_2$ conditions showed a moderate enhancement of PL in the first 1-2 minutes, followed by a mild decline of PL intensity. On the other hand, a sample that has been exposed to oxygen and light for 15 minutes prior to changing the environment to N$_2$ showed a drastic decline in PL. Interestingly, if this preconditioning was performed in air rather than O$_2$, changing the environment to N$_2$ resulted in an increase in PL intensity, as was observed for pristine samples. This observation is interesting since it suggests that not only PL characterization in vacuum may be influenced by desorption of physisorbed molecules (e.g. O$_2$) from the sample surface, but possibly also measurements in $N_2$. Another early report by Gottesman *et al.* investigated the PL evolution of MAPbI$_3$ samples sealed in $N_2$ and showed that the PL strongly decreases upon illumination, which can be reversed by storing the samples in the dark.[300] The authors assigned this behavior to a reversible structural transformation of the MAPbI$_3$ crystal under illumination.

Following these initial reports, both an increase and decrease of PL under $N_2$ conditions have been reported in literature. Many groups observed an enhancement of PL under $N_2$ and illumination,[41,301,302] leading to the term 'photo-brightening'. Seminal work by Stranks and coworkers tied this effect to ion migration that leads to trap annihilation in illuminated areas.[231] However, a number of studies also reported a decline of PL in $N_2$ environments. For example, using PL microscopy on polycrystalline MAPbI$_3$, Brenes *et al.* demonstrated that both initially dark and bright grains show a decrease in PL upon exposure to $N_2$.[303] On the other hand, Hong *et al.* reported that both PL enhancement and decline under $N_2$ can occur on the same timescale of microseconds to seconds even on different areas of the same MAPbI$_3$ microcrystal.[304] The authors proposed that it is the relative concentration and quenching efficiency of brightening and darkening traps that is responsible for these opposing behaviors. Supported by theoretical calculations, they postulate that $I_i^{+1}$ defects may be responsible for PL enhancement, while $Pb_i^{2+}$ for decline of PL. In another example, while MAPbI$_3$ polycrystalline deposits showed an increase in PL in $N_2$, MAPbBr$_3$ showed a significant decrease.[305] Similar to the case of characterization under vacuum, it is likely that an intricate interplay between ion migration, defect formation and annihilation and other light-induced processes determines both the trend and dynamics of PL evolution under $N_2$ conditions, with significantly more research required to fully elucidate these processes and their exact contributions.

*6.2.3. Oxygen*

The effect of oxygen on the PL of perovskite materials gained significant interest from the research community following the 2015 report by Tian *et al.* demonstrating a thousand-fold enhancement in PLQE of MAPbI$_3$ following exposure to light and O$_2$.[306] Similar enhancements were also reported for other perovskite compositions and structures, such as MAPbBr$_3$ single crystals,[278] thin films of triple cation perovskites[307] and CsPbBr$_3$ nanowire ensembles[308] and nanocrystals.[309]

A detailed investigation by Anaya *et al.* revealed that exposure to oxygen can lead to an enhancement in the PL of MAPbBr$_3$ even without illumination.[310] This experimental observation is supported by a theoretical study by Meggiolaro *et al.* which suggests that oxygen can deactivate deep hole traps associated with iodide interstitials.[311] Anaya *et al.* also demonstrated that the concentration of oxygen in the environment plays a role in determining the dynamics and range of PL intensity evolution. By combining PL studies with compositional studies using X-ray photoemission spectroscopy, the authors proposed the following three stages in the interaction of O$_2$ with the perovskite: (1) shortly after photoexcitation, electrons are captured by molecular oxygen to form superoxide (O$_2^-$) at the surface of the perovskite sample, while holes are captured by halide anions to form halide interstitials. No significant changes in the PL occur at this initial stage. (2) After the density of surface O$_2^-$ has risen to sufficiently high levels, the interstitial halides begin to migrate towards the bulk of the perovskite sample, annihilating halide vacancies and reducing the non-radiative trap density, thus enhancing the PL intensity. (3) Finally, at longer times, the PL is reduced following degradation of the perovskite sample via previously reported routes.[22,312]

Unlike the mechanism proposed by Anaya *et al.* Islam and co-workers suggest that superoxide species can directly occupy iodide vacancy cites at the surface and bulk of the perovskite layer due to O$_2^-$ being of similar size to the iodide ion.[313] Indeed, fast diffusion of O$_2$ into the bulk of perovskites layers has been shown to occur in the dark[310] and to be further accelerated upon illumination,[314] and enhanced PL at the bulk of single crystal MAPbI$_3$ exposed to O$_2$ has been reported by Feng *et al.*[315] This mechanism is further supported by PL microscopy studies which show that initially dark grains, in which the density of such defects is high, brighten upon exposure to $O_2$, while bright grains do not.[301] Recent theoretical calculations by Prezhdo and co-workers also strongly support this mechanism, demonstrating that both superoxide and peroxide can passivate halide vacancies.[316]

While most reports show that exposure to $O_2$ results in an increase in PL intensity, it has been recently shown that a decrease in PL may also be observed. For example, Fassl *et al.* investigated a range of MAPbI$_3$ samples with fractionally varying stoichiometries and demonstrated that while understoichiometric samples showed an enhancement in emission intensity, even slight overstoichiometry caused a decline (Figure 17c).[180] Strong decline of PL upon exposure to $O_2$ has also been observed by Mantulnikovs *et al.*[305] Durrant and co-workers recently suggested that such a decrease in PL upon exposure to oxygen may occur in samples in which surface defects are well passivated.[317] They postulate that in such samples, oxygen present at the grain boundaries which is not incorporated into perovskite defects forms the superoxide species by electron capture, thus reducing the PL intensity. These studies highlight the need for further research into the mechanisms by which oxygen interacts with perovskite defects and its effect on perovskite PL.

*6.2.4. Humidity*

The study of the effect of moisture on the properties of perovskites was triggered by the observations of severe water induced degradation in these materials.[318–320] Early work by Grancini *et al.* showed that water can intercalate into the edges of MAPbI$_3$ single crystals, resulting in widening of the bandgap (i.e. blue shift of the PL spectrum) and shortening of PL lifetime.[297] This observation was supported by Müller *et al.* which showed that water molecules can infiltrate perovskite films on a timescale of seconds.[321] On the other hand, a moderate enhancement in PL of single crystal MAPbBr$_3$ upon exposure to moisture has been observed by Fang *et al.* and assigned to passivation of surface trap states by water molecules.[293] The same authors also reported a remarkable ~150 times enhancement of PL in FAPbI$_3$ thin films upon exposure to humidity, which was assigned to defect healing.[322] In the case of mixed halide I$_x$Br$_{1-x}$ perovskites, the effect of humidity on the PL was shown to be strongly influenced by the composition of the samples, possibly due to the effect of water intercalation on ion transport in such materials.[323] While a similar enhancement in PL upon exposure to humidity has been reported by Brenes *et al.*, their theoretical calculations suggest that water molecules do not interact with surface defects as occurs upon exposure to molecular oxygen.[301] The authors suggest that instead exposure to moisture forms a thin amorphous shell of PbI$_2$ or PbO, which eliminates surface defects, resulting in a largely irreversible increase in PL.

The PL enhancing effects of exposure to O$_2$ and humidity seem to be cumulative. Brenes *et al.* showed that upon exposure to humid air, both the PL intensity and lifetime of MAPbI$_3$ is increased beyond what was observed for O$_2$ or humidity alone.[301] Similar observations were also reported by Fang *et al.* on MAPbBr$_3$ and by Dänekamp *et al.* for vacuum processed MAPbI$_3$.[273,293] However, while the overall PL enhancement upon exposure to humid air is large, the presence of humidity may lead to a subsequent decay in PL on a timescale of minutes.[295,298] This highlights that PL measurements performed in humid air may exhibit photodarkening due to accelerated degradation in O$_2$ and moisture.[324]

## 6.3. Reporting Experimental Conditions

While the choice of excitation parameters and experimental environment are critically important when performing PL characterization of perovskites, a number of other influences on PL have been reported in literature. For example, storing perovskite samples prior to PL characterization has been shown to either increase[325] or decrease[326] PL lifetimes in perovskite films. While the vast majority of PL characterization experiments are performed on perovskite films deposited on glass, it has been shown that migration of Na$^+$ ions from glass substrates into the perovskite layer may significantly enhance PL lifetimes if stored for a day prior to PL measurement.[299] Furthermore, although some groups report the use of protective layers such as PMMA[300] or Cytop[301] to prevent perovskite samples from degrading during PL characterization studies, such layers do not fully suppress diffusion of neither O$_2$ nor water, and may on their own influence the overall PL behavior.

The range of influencing experimental factors is so broad that accurate reporting of such factors becomes a critical issue. In **Figure 19**, we summarize eight of the most critical factors that may influence the results of PL measurements. With the abundance of contradicting reports in literature, precise reporting of experimental conditions is the first step towards increasing the reproducibility of results in the field of perovskite materials. We highly encourage all researchers to consider these factors upon designing and executing their experiments and to systematically report these parameters in publication of their results.

Unless the experiment is directly related to the study of environmental influences, we recommend that whenever possible, researchers perform PL measurements in inert environments such as N$_2$ or Ar in order to limit the impact of environmental factors. We also suggest that instead of a single PL spectrum and/or PLQE value obtained at some point during the PL experiment, it becomes routine to show the evolution of these spectra/values as a function of time for 10 minutes from the start of the experiment. To further account for the impact of the sample history, we recommend that researchers doing a series of measurements check that the values obtained are actually independent of each other when they are supposed to be. For example, when evaluating the impact of laser power on PLQE by incrementally increasing the laser power on the same sample spot, one has to account for the fact that illumination time is also a variable that impacts PLQE, and the two cannot be easily decoupled. In fact, it should become routine to characterize several areas of the sample in order to evaluate how representative a certain measurement is of the overall sample. Finally, due to the impact of excitation power, we advise the use of excitation fluence equivalent to 1 sun for simple characterization of the PL, which is not only comparable to operation conditions of a solar cell but will also allow for a better comparison of values obtained by different groups.

## 7. Outlook and Conclusions

In summary, we described the current state of the observation of photoluminescence in perovskites for optoelectronic devices. Photoluminescence is important not only for an increased understanding of the fundamental photophysical properties of these high-performance semiconductors, but also because it can be used as a tool to engineer higher performing devices. This is in part due to its relationship to the open-circuit voltage of the solar cell – any means to maximize luminescence efficiency through a reduction in non-radiative recombination pathways will also result in an increase in the $V_{OC}$. In accomplishing this task, linking enhancements in the PLQE to specific changes in the complete composition of the perovskite film – bulk crystal structure, defects, and microstructure – becomes vital. The experimental effort in doing so has been tremendous, with observations of how the recipe environment, excess or deficient halide content, and non-uniformity in the microstructure all impact the observed luminescence. Surface treatments have proven especially effective at passivating defects and enhancing both PL and device performance. Unfortunately, while there is widespread usage of the relative intensity of the PL of different perovskite compositions within the same experiment, PLQE is not widely used, preventing adequate comparison between groups. Most critically, when PLQE is reported, we observed a preponderance of values measured for the same type of film for the same excitation density, suggesting hidden variables which impact the defect content of the devices. These are partially accounted for by natural sample-to-sample variations which occur when moving between labs, including slight changes to recipes resulting in subtle changes to the defect content, and changes in light outcoupling between samples due to variable surface roughness. They especially come to light when noting the (sometimes opposing) trends for behavior of the photoluminescence under varying measurement conditions. The environment – oxygen content, humidity – the sample history and storage conditions, any protective coatings, the excitation power, and more, are all shown to have an impact on a sample's maximum PLQE as well as its behavior over time. Based on these observations, we propose that it would be mutually beneficial to all interested research groups to adopt a common set of measurement conditions when reporting the simple PLQE for a perovskite film. Our recommendation is that measurements be performed in an inert atmosphere, that the value be tracked for 10 minutes, that the excitation power be set to 1 sun in intensity, and that multiple spots be checked for consistency. Such practices will aid researchers in coming to a consensus on the underlying photophysical properties of perovskites, especially with regards to the role that defects do or do not play in the ultimate behavior of a film or device.

**Acknowledgements**

The authors acknowledge funding from the European Research Council (ERC) under the European Union's Horizon 2020 research and innovation programme (ERC Grant Agreement n° 714067, ENERGYMAPS). The authors also acknowledge funding from Deutsche Forschungsgemeinschaft (DFG) SPP 2196, project VA 991/3-1.

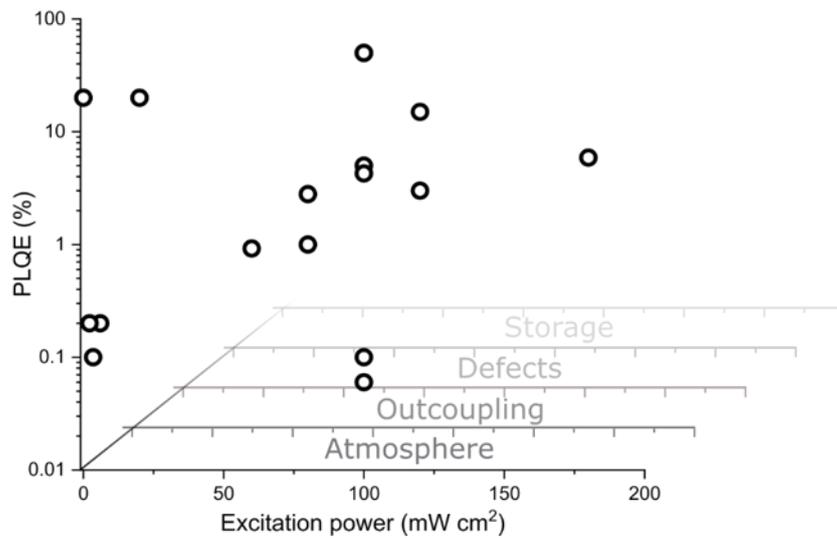

**Figure 1**: Selection of PLQE values reported in the literature as a function of excitation power for MAPbI$_3$ without modifications. There is no discernible trend, illustrating the sensitivity of the PLQE measurement to "hidden" variables, such as different outcoupling coefficients, unintentional defect introduction, or storage conditions, and emphasizing the need for a more standardized measurement to facilitate comparable PLQE measurements between research groups. Values taken from the references [37-48].

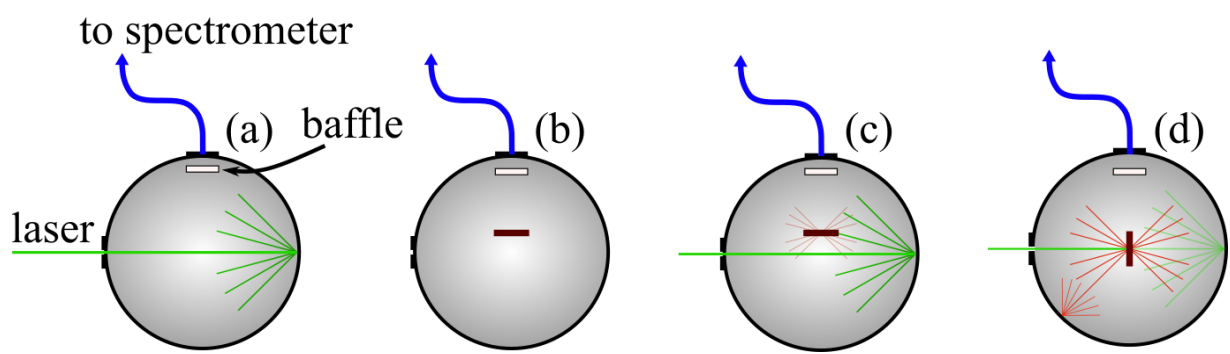

**Figure 2**: Schematic of the integrating sphere's interior during measurement of the PLQE. A baffle is placed immediately in front of the optical fiber to prevent any direct illumination from the laser. (a) Reference laser intensity. (b) Background signal. (c) Measurement with sample out of beam path to measure indirect excitation. (d) Measurement with sample inline to measure direct photoluminescence signal. Adapted with permission from reference [35]. Copyright 2004, Wiley.

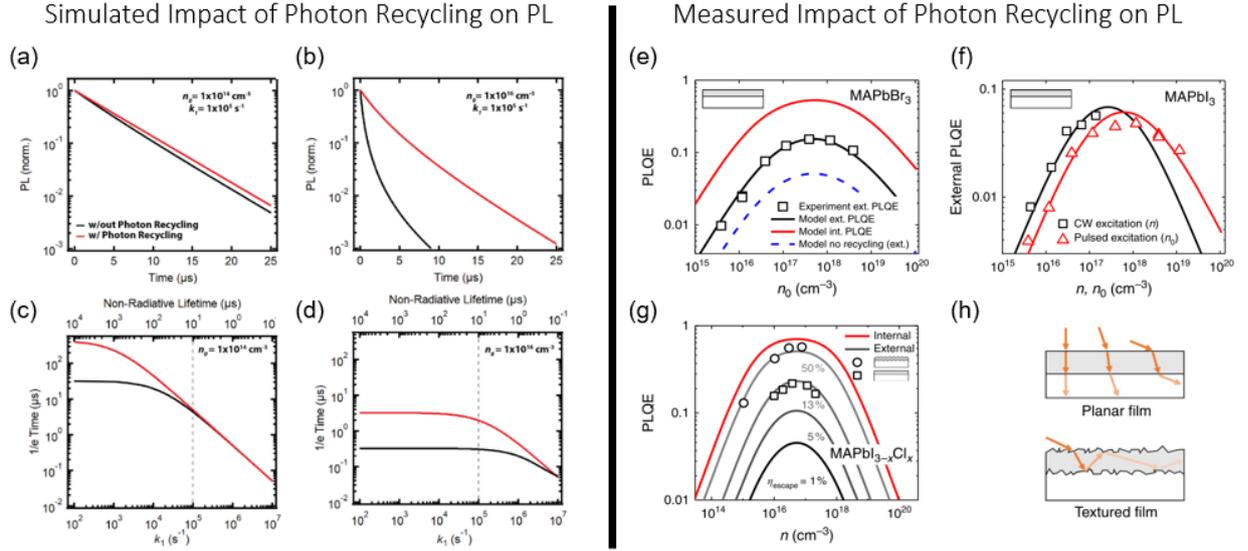

**Figure 3**: PLQE dependence on photon outcoupling and fluence dependency. Simulated TRPL curves at (a) low and (b) high fluences for fixed nonradiative and radiative recombination coefficients $k_1$ and $k_2$, with and without considering photon recycling effects. (c), (d) Time required for PL to decay to 1/e of its initial value, for a range of $k_1$ values, again for (c) low and (d) high fluences. Dotted lines indicate the $k_1$ values plotted in (a,b). (e) Measured $\eta_{ext}$ (squares), modelled $\eta_{int}$ (red curve) and $\eta_{ext}$ (black curve), and modelled $\eta_{ext}$ (blue dotted line) without photon recycling for MAPbBr$_3$ thin films. (f) Modelled (curves) and measured (squares, triangles) $\eta_{ext}$ for MAPbI$_3$ thin films for pulsed and continuous wave excitation. (g) Measured $\eta_{ext}$ for MAPbI$_{3-x}$Cl$_x$ thin films fabricated on textured (circles) and planar (squares) substrates, together with modelled curves for various photon escape probabilities. (h) sketch illustrating the superior incoupling of the textured film as compared to regular planar architecture. (a-d) Adapted with permission from reference [56]. Copyright 2019, American Chemical Society. (e-h) Adapted from reference [57] under the terms of the Creative Commons CC BY License. Copyright 2016, Springer Nature.

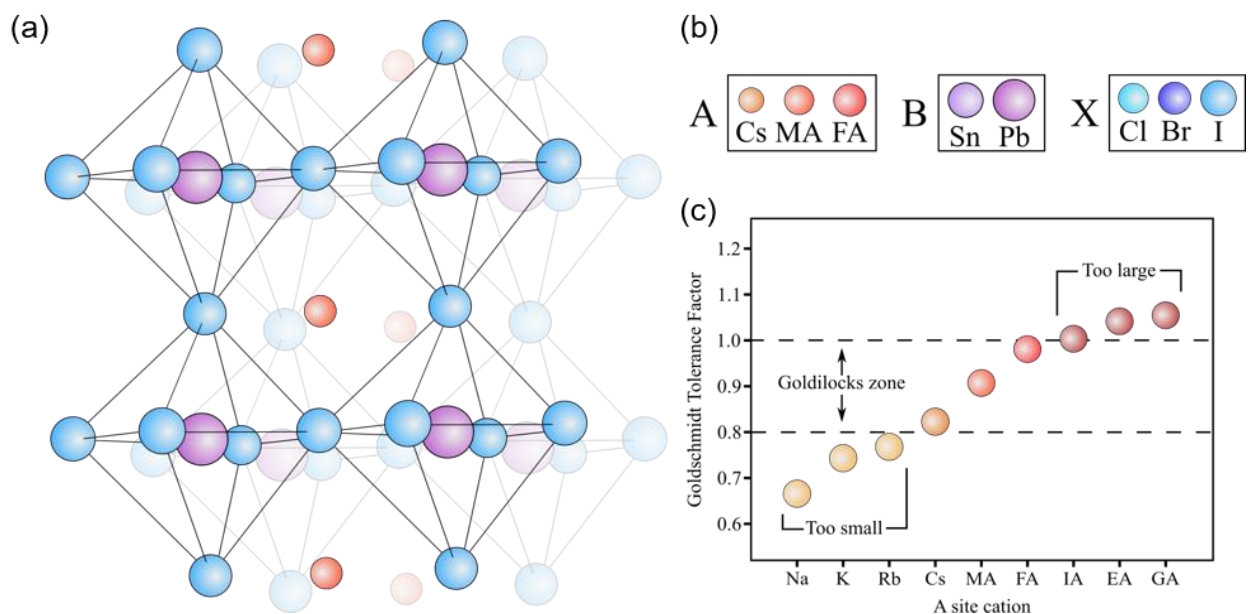

**Figure 4**: Illustration (a) of the perovskite ABX$_3$ tetragonal crystal structure, with a short list (b) of commonly employed elements for each site. (c) The Goldschmidt tolerance factor states that only Cs, MA, and FA can form perovskite, as their ionic radius is "just right," while other molecules are either too small or too big. Adapted with permission from reference [90], Copyright 2017, American Association for the Advancement of Science.

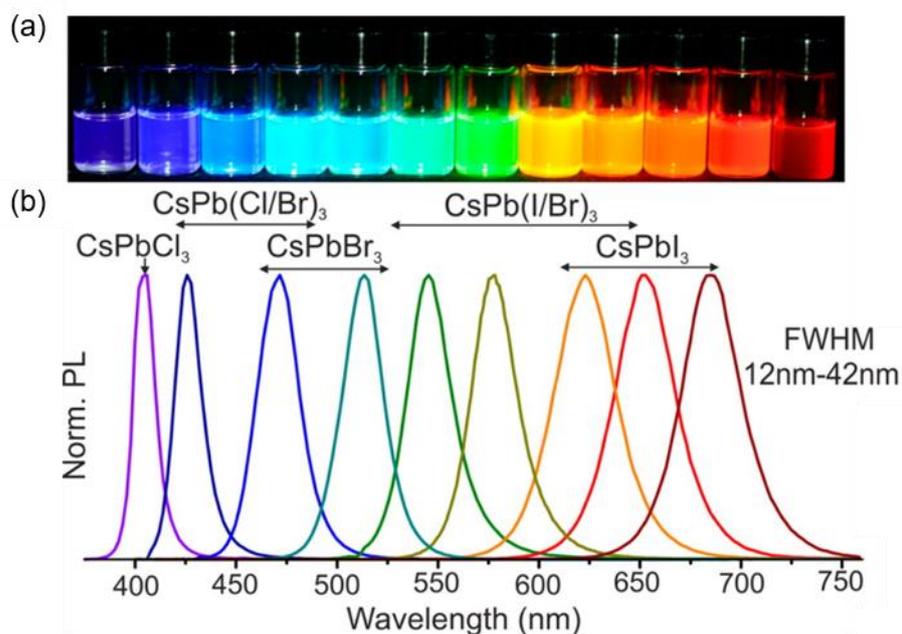

**Figure 5**: (a) Solutions of CsPbX$_3$ perovskite quantum dots can emit any color in the visible spectrum via halide substitution, using combinations of X = I, Br, Cl. (b) Normalized PL spectra showing narrow emission for every color. Both adapted with permission from reference [124], Copyright 2015, American Chemical Society. Further permissions related to this can be found at https://pubs.acs.org/doi/10.1021/nl5048779.

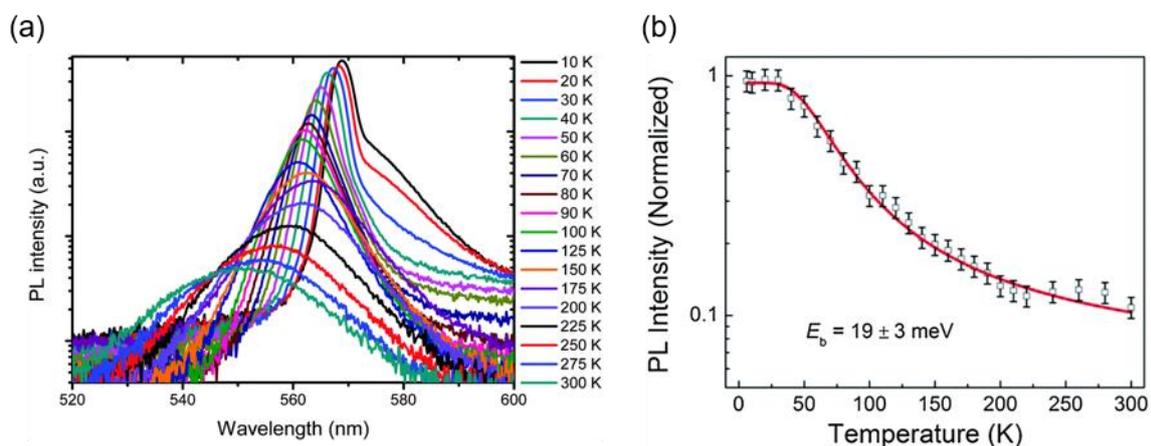

**Figure 6**: (a) PL spectra of FAPbBr$_3$ microstructures measured at different temperatures Adapted with permission from reference [146], Copyright 2013, Royal Society of Chemistry. (b) Decay of integrated PL intensity with temperature of a CH$_3$NH$_3$PbI$_3$ film under excitation with a continuous-wave 532 nm laser. The solid line is the fit based on the Arrhenius equation used to describe the observed functionality. Adapted from reference [144] under the terms of the Creative Commons Attribution, Non-Commercial 3.0, unported license - Published by the Royal Society of Chemistry.

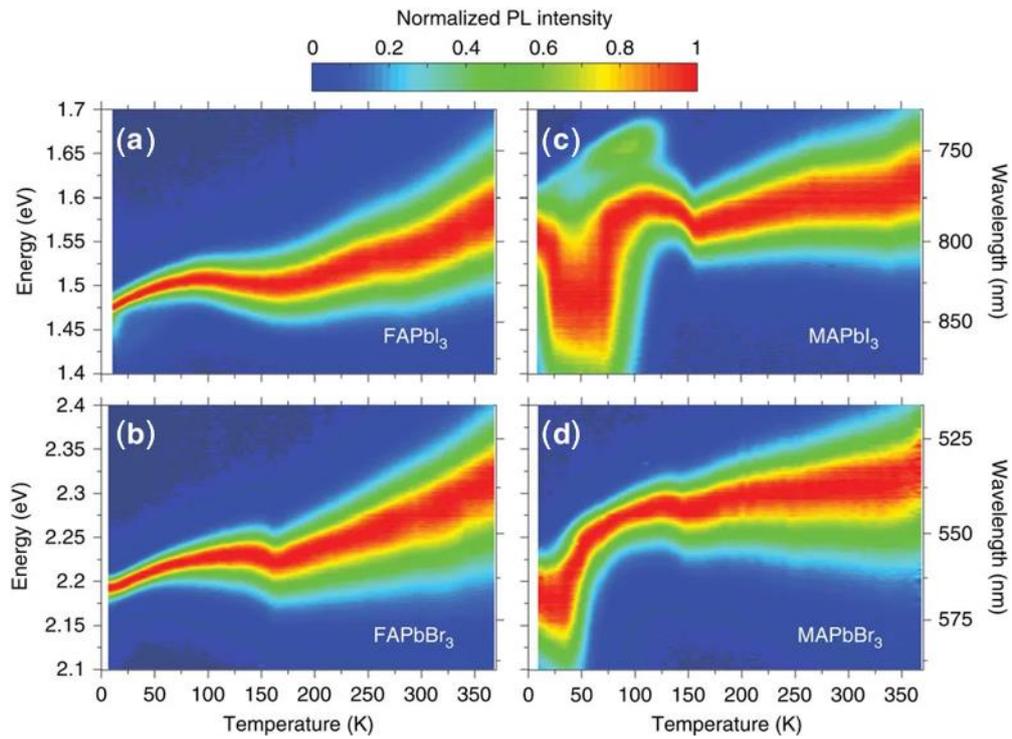

**Figure 7**: Temperature dependence of normalized steady-state PL between 10 and 370 K of (a) FAPbI$_3$,(b) FAPbBr$_3$,(c) MAPbI$_3$ and (d) MAPbBr$_3$ thin films. Adopted from reference [157] under the terms of the Creative Commons Attribution 4.0 International License, Copyright 2016, Springer Nature.

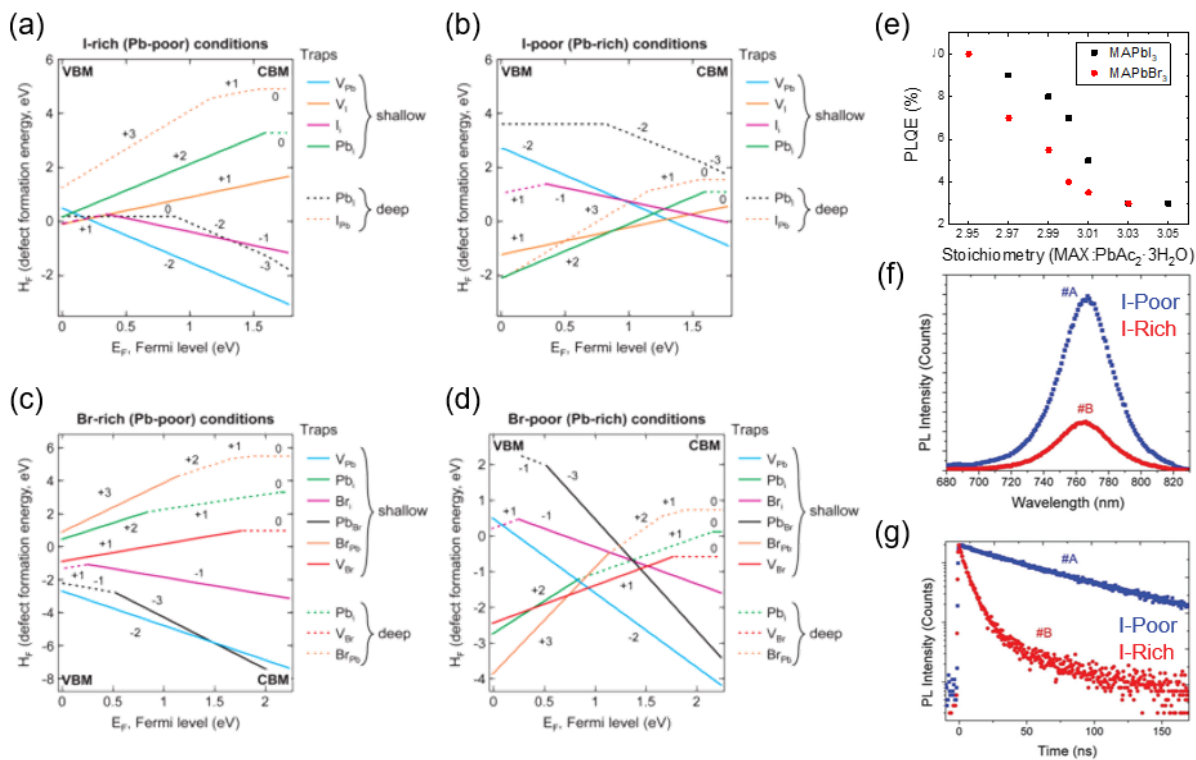

**Figure 8**: Calculated defect formation energies for MAPbI3 in (a) I-rich and (b) I–poor environments. Calculated defect formation energies for MAPbBr3 in (c) Br-rich and (d) Br-poor environments. (a-d) Adapted with permission from reference [175]. Copyright 2015, American Chemical Society (e) PLQE as a function of precursor stoichiometry for MAPbI$_3$ and MAPbBr$_3$. Adapted from references [180,181]. [180]-reproduced by permission of the Royal Society of Chemistry. [181]-reproduced under the terms of the Creative Commons CCBY License, Copyright 2019, Wiley. (f) PL intensity comparison of two differently processed samples with different iodide content: #A being iodide poor and #B iodide rich. (g) I-rich and I-poor TRPL traces, measured at peak intensity wavelength. (f, g) adapted with permission from reference [186]. Copyright 2016, Royal Society of Chemistry.

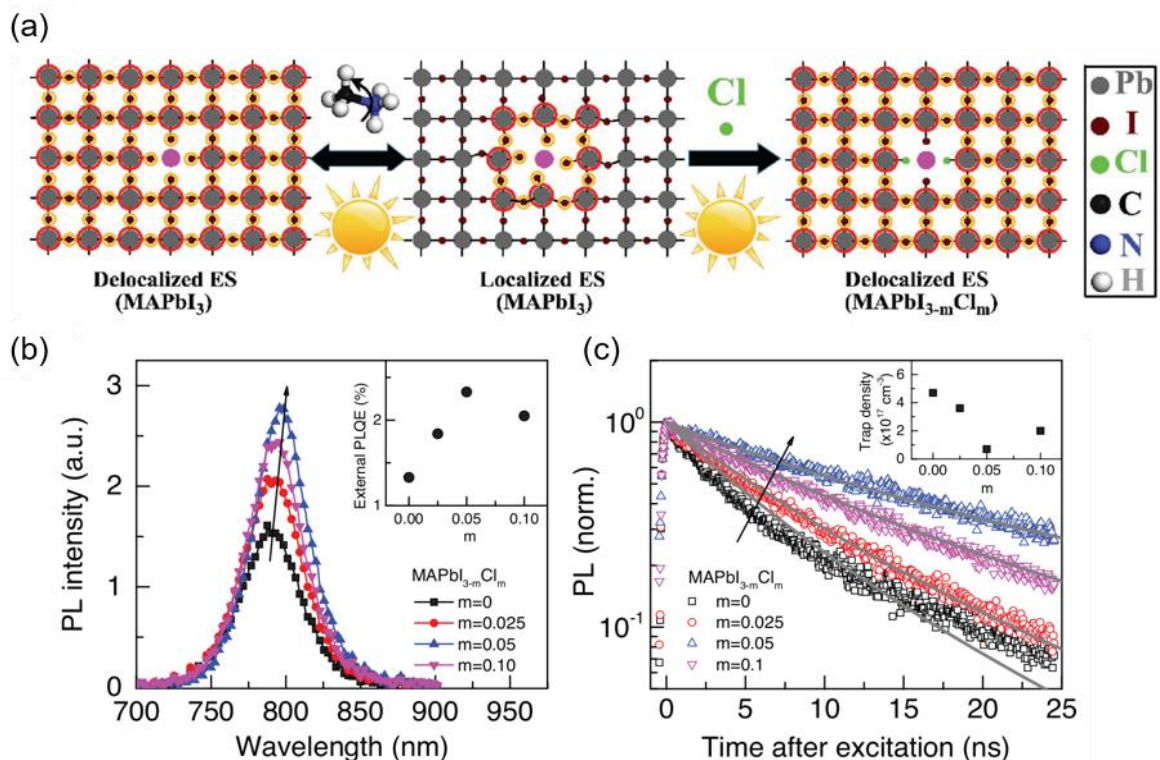

**Figure 9**: Schematic detailing the mechanism of healing in MAPbI$_3$ and MAPbI$_{3-m}$Cl$_m$, either dynamically via MA reorientation or statically via Cl doping. The vacancy is the magenta dot, the electron density as red circles around the grey lead atoms, and the hole density as yellow circles around the brown iodide atoms. The (de)localization of the excited states (ES) is thus represented by the extent of these red and yellow circles. (b) Steady state PL (532 nm illumination, CW) increases with Cl doping to an extent, before beginning to decrease. The inset shows the external PLQE values averaged across three points per sample, which maximize at 2.3%. (c) The corresponding TRPL measurements (407 nm illumination, 3.2uJ cm$^{-2}$ fluence, pulsed) show an increase in the PL lifetime. Grey lines are the fits calculated from a trap model, and the resulting trap densities from these fits are displayed in the inset. Adapted with permission from reference [182], Copyright 2019, Wiley.

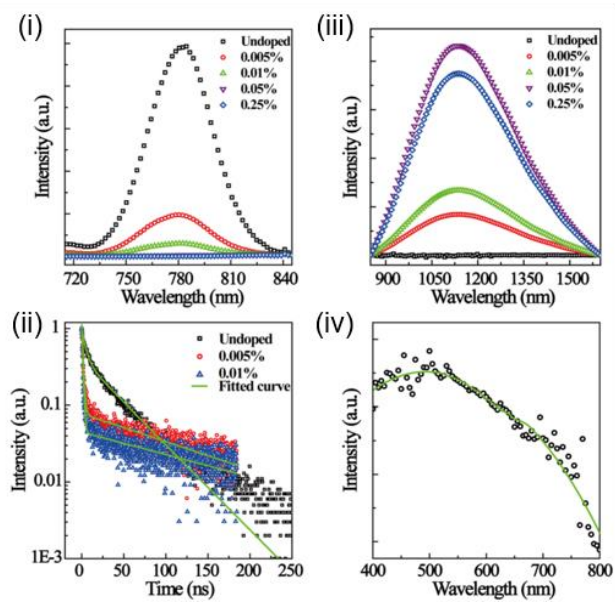
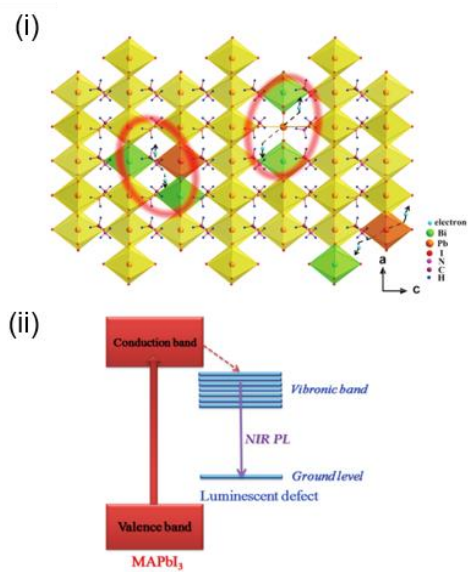
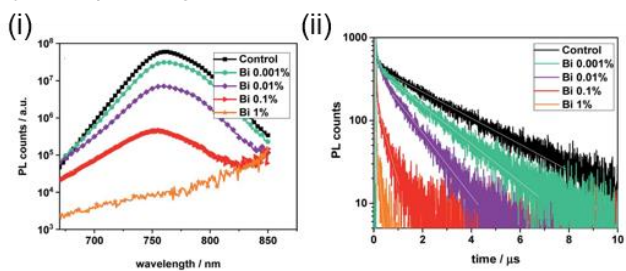
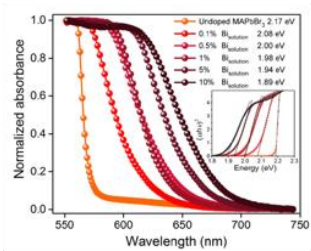

**Figure 10**: Comparison of PL spectra (517 nm illumination) of pristine and Bi-doped MAPbI$_3$ films, in the (i) visible and (iii) IR spectral regions. (ii) TRPL measured at 782 nm for the same films, with the fitting lines shown in green. (iv) Excitation intensity of the 0.01% doped film, measured at 1150 nm, indicating that the photoexcitation wavelength extends to 780 nm. The green line is shown to help guide the eye. (b) Schematic illustration of the origin of NIR PL. (i) Crystal structure of Bi-doped MAPbI$_3$, with the proposed NIR luminescent centers circled in red. (ii) Energy diagram of Bi-doped MAPbI$_3$. After photoexcitation (solid red arrow), the MAPbI$_3$ excitation is transferred to a lower energy gap defect state (red dotted arrow), leading to the observed NIR emission (violet arrow). (a,b) Adapted with permission from reference [195]. Copyright 2016, American Chemical Society. (c) (i) Steady state and (ii) TR PL spectra (460 nm illumination, pulsed at 50 kHz, 11 pJ pulse$^{-1}$) of Bi-doped CsFAMA films. Doping decreases both the absolute intensity as well as the lifetime, to the point at which it is barely noticeable by 1% doping. Adapted with permission from reference [197]. Copyright 2013, Royal Society of Chemistry. (d) Steady state absorbance of Bi-doped MAPbBr$_3$ single crystals. The inset displays the corresponding Tauc plots. Adapted with permission from reference [196]. Copyright 2016, American Chemical Society.

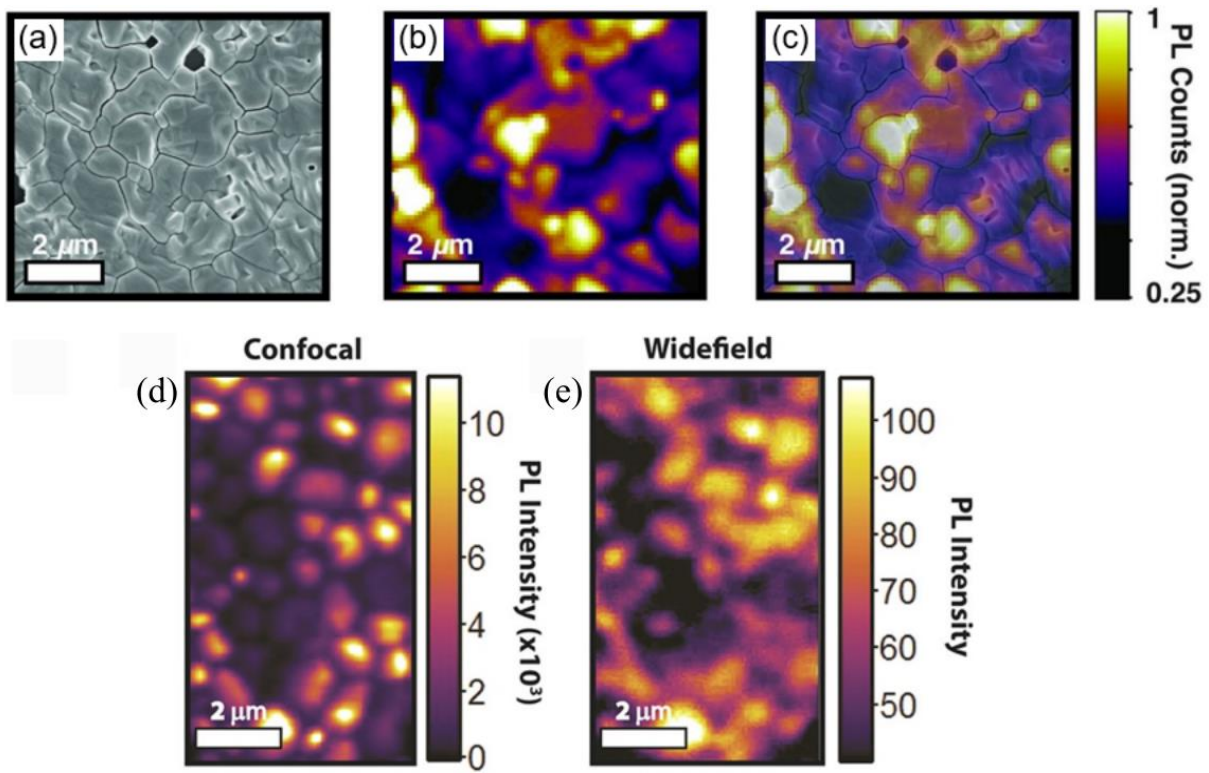

**Figure 11**: (a) SEM micrograph and (b) fluorescence image (470 nm illumination, 30 nJ cm$^{-2}$, 125 kHz pulsed) of the same area, and (c) composite of both, showing significant spatial variations in PL intensity of an MAPbI$_{3-x}$Cl$_x$ film. Adapted with permission from [59]. Copyright 2015, American Association for the Advancement of Science. (d) Confocal (470 nm illumination, 0.2 μJ cm$^{-2}$, 2.5 MHz) and (e) wide-field (532 nm illumination, 220 mW cm$^{-2}$, CW) fluorescence images of the same area, confirming there is no significant difference in heterogeneity in the observed PL spectra. Adapted with permission from reference [211]. Copyright 2017, American Chemical Society.

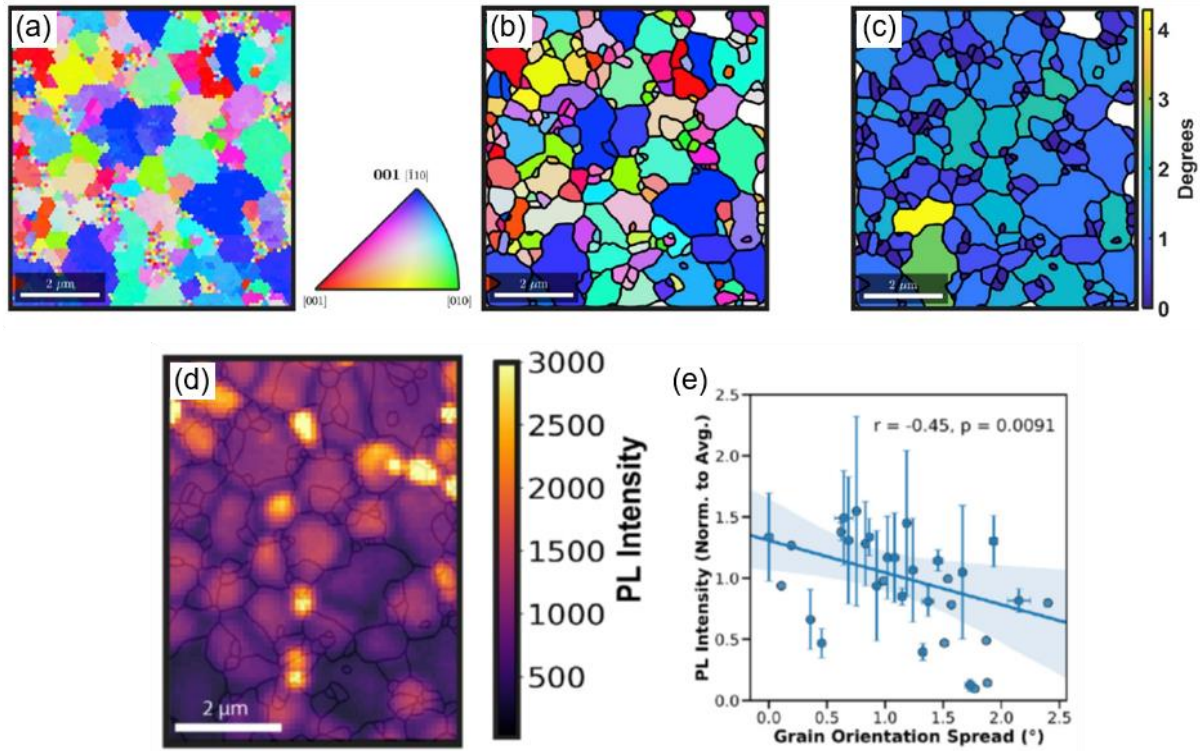

**Figure 12**: Results of EBSD experiments on MAPbI$_3$ thin films. (a) IPF map generated from IPF color key to identify different orientations. (b) Grain structure found using an orientation threshold of 4°, and colored according to their mean orientation. (c) Grain orientation spread (GOS) from (b), plotted with the mean local misorientation showing neighboring grain heterogeneity. (d) Composite of grain network obtained from EBSD measurements with confocal PL (470 nm illumination, 0.2μJ cm$^{-2}$, 1 MHz). PL intensity as a function of GOS, showing strong negative correlation (p-value = 0.0091). The solid line is a linear fit to the data, with the shaded region representing a 95% confidence interval for the line of best fit. Adapted with permission from reference [206]. Copyright 2019, Elsevier.

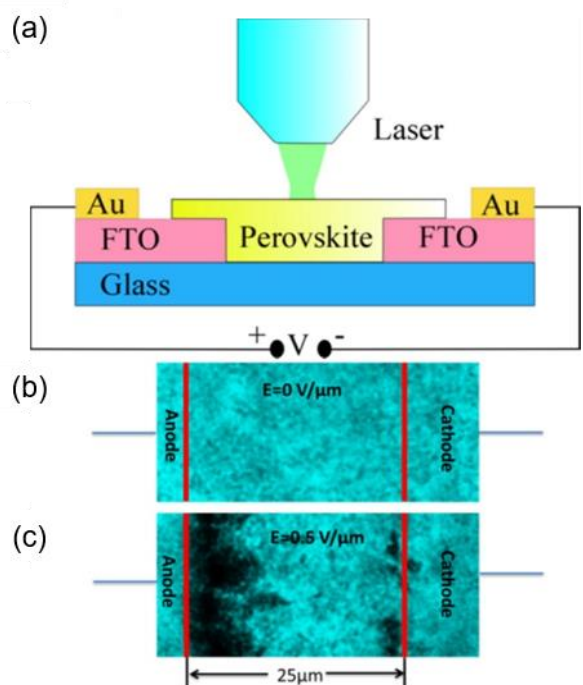

**Figure 13**: (a) Schematic of electric field-dependent PL microscopy measurement. Optical images (b) with and (c) without applied lateral bias, showing ion diffusion response. Adapted with permission from reference [230]. Copyright 2016, American Chemical Society.

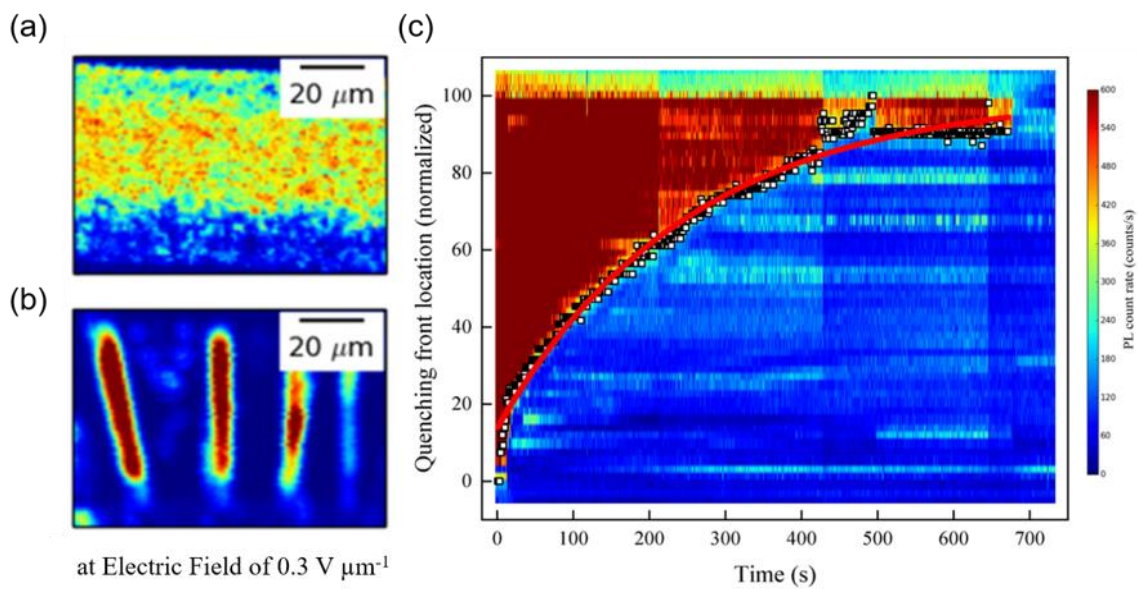

**Figure 14**: Spatial PL maps of MAPbI$_3$ (a) spin-coated or (b) zone-cast under an applied bias of 0.3 V μm$^{-1}$. The PL signal is partially quenched from the migration of negative (positive) II-/VMA- ions toward the bottom (VI+/MAi+ toward the top) of the channel. (c) Time evolution of the quenching front under a bias of 0.9 V μm$^{-1}$ of a single zone-cast ribbon. The solid red line is the line of best fit obtained assuming the ions follow a linear diffusion model v = μE, and that the accumulation of ions screens the applied electric field E, causing an exponential decrease with respect to time. Adapted with permission from reference [258]. Copyright 2019, American Chemical Society.

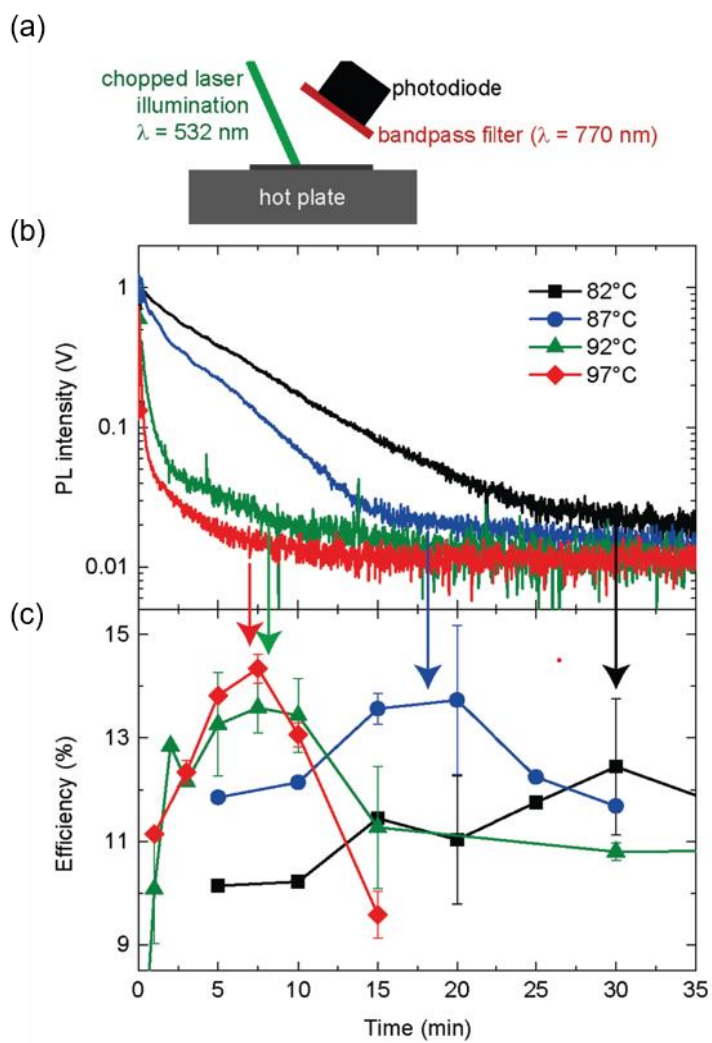

**Figure 15**: *In situ* steady state PL (532 nm illumination, 0.09 MW, 534 Hz) intensity of annealing MAPbI3 thin films. a) The schematic representing the *in situ* measurement setup b) PL intensity for various annealing temperatures, showing different rates of increasing quenching. c) The corresponding PV performance for devices annealed for the time and temperature indicated by the arrow. Adapted with permission from reference [32]. Copyright 2016, Wiley.

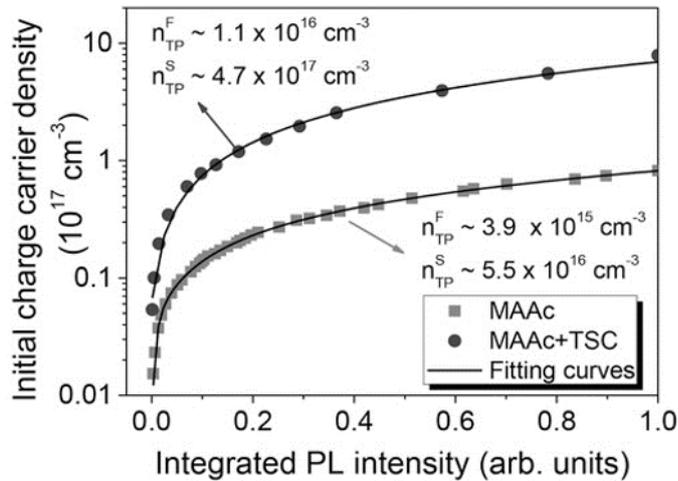

**Figure 16**: Temporally integrated PL (460 nm illumination, 0.01-2 μJ cm$^{-2}$, 80 MHz) vs. charge carrier density (TIPL) measurement for MAPbI$_3$ films with MAAc additive (dark circles) and MAAc + TSC additive (grey squares). Fitting is according to equation 16 assuming densities of slow and fast trapping defects, with the corresponding trap density values listed. Note that the legend in the original publication appears to be mislabeled. Adopted with permission from reference [274]. Copyright 2017, Wiley.

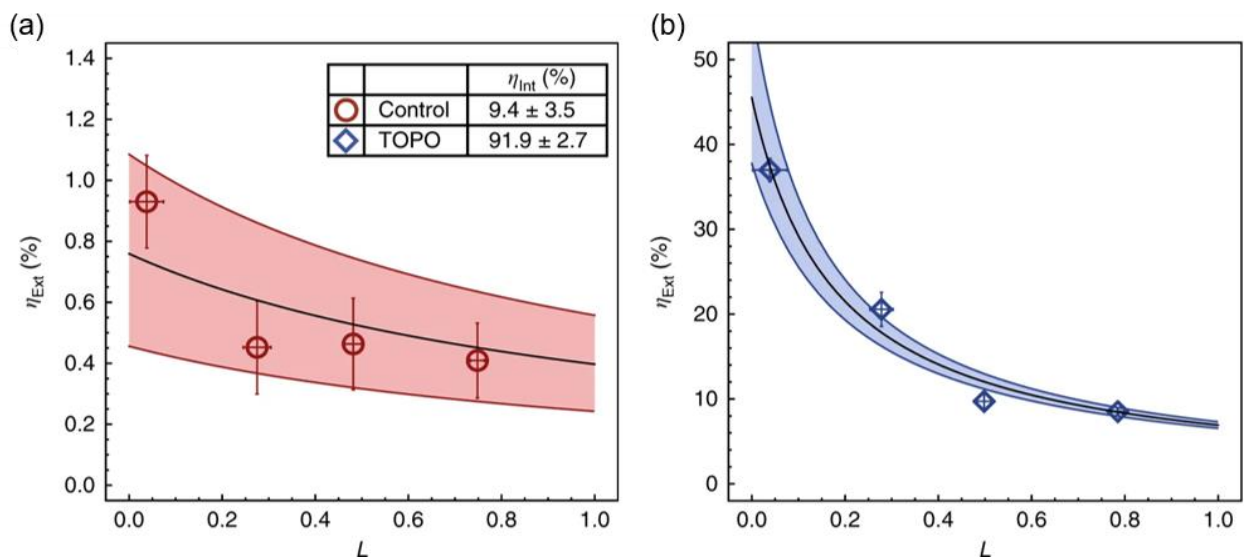

**Figure 17**: Determination of the external PLQE (540 nm illumination) for MAPbI$_3$ films by varying back-surface parasitic absorption (*L*), (a) without and (b) with a TOPO passivation treatment. Shaded regions represent a 95% confidence interval of the black line of best fit. Adopted with permission from reference [37]. Copyright 2018, Springer Nature.

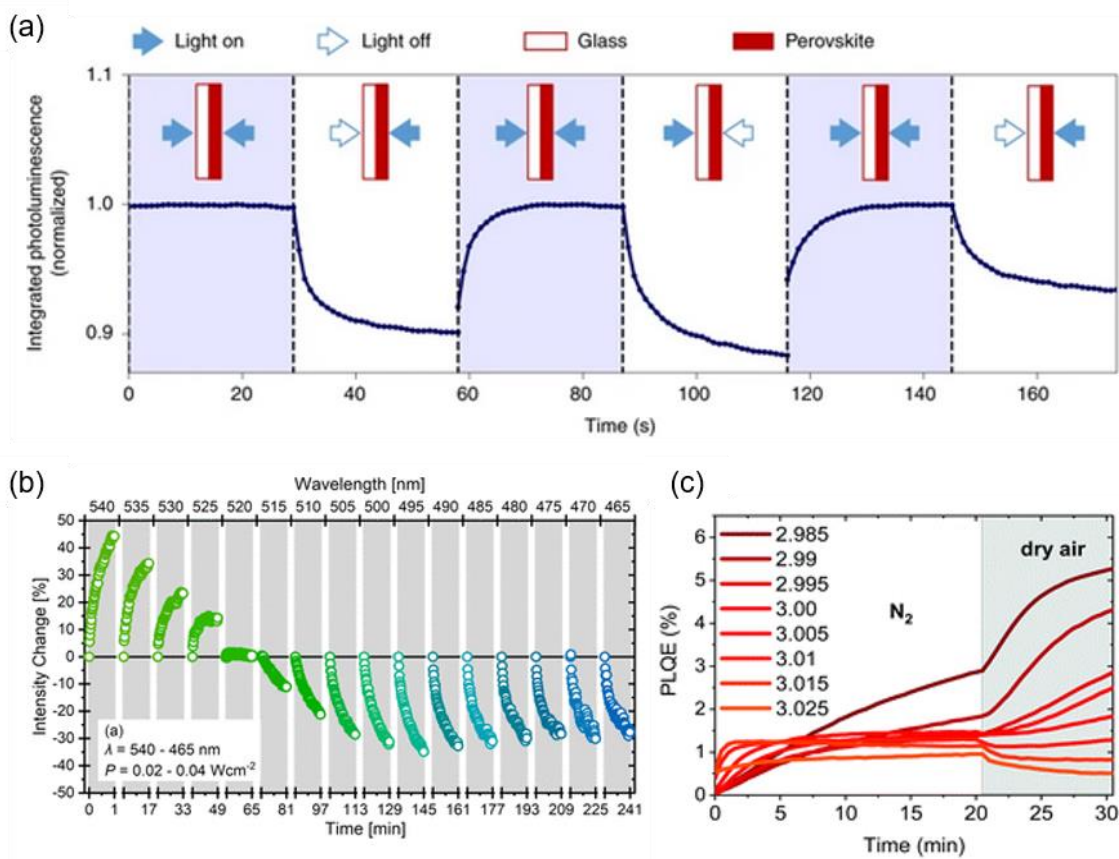

**Figure 18**: Examples of opposing trends in evolution of PL for MAPI$_3$ films. (a) Integrated PL intensity (560 nm illumination) as a function of time of, with illumination striking the sample from either the perovskite or glass side. The two beam paths were adjusted to balance intensity, and for each configuration the trace is normalized to its maximum intensity for easier visualization. Adapted with permission from reference [288]. Copyright 2019, Springer Nature. (b) Steady state PL traces for decreasing illumination wavelength measured at measured at the PL peak position. The samples were illuminated for 60 s at each wavelength, and then kept in the dark for 15 min to reduce sample stress. Adapted with permission from reference [289]. Copyright 2018, American Chemical Society. (c) Time evolution of the PLQE for MAPbI$_3$ films fabricated with varying stoichiometries (440 nm illumination, 80 mW cm$_{-2}$, CW), inside an integrating sphere flushed with N$_2$ and dry air. Adapted under the terms of the Creative Commons Attribution, non-commercial 3.0, unported license. From reference [41] – Published by the Royal Society of Chemistry.

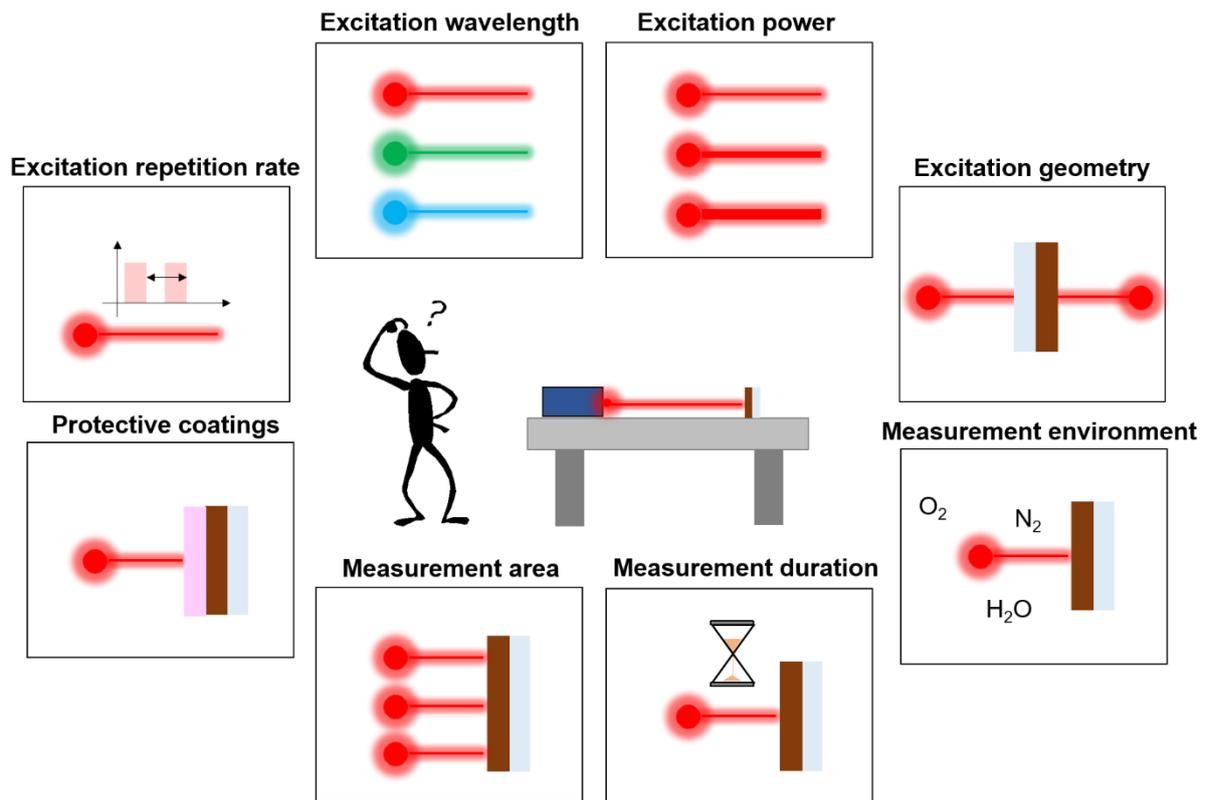

**Figure 19**: Summary of the experimental conditions that can influence the measurement of PL in perovskite materials.